\documentclass[pdflatex, sn-basic]{sn-jnl}

\usepackage{graphicx}
\usepackage{multirow}
\usepackage{amsmath,amssymb,amsfonts}
\usepackage{amsthm}
\usepackage{mathrsfs}
\usepackage[title]{appendix}
\usepackage{xcolor}
\usepackage{textcomp}
\usepackage{manyfoot}
\usepackage{booktabs}
\usepackage{algorithm}
\usepackage{algorithmicx}
\usepackage{algpseudocode}
\usepackage{listings}
\usepackage{array}

\usepackage{tikz}
\usetikzlibrary{quantikz2}
\usepackage{bbm}
\usepackage{xspace}
\usepackage{placeins}
\usepackage{caption}
\usepackage{makecell}
\usepackage{enumitem}

\newcommand\abs[1]{\left\lvert#1\right\rvert}

\newcommand{\eg}{e.g.,\xspace}
\newcommand{\idest}{i.e.,\xspace}

\newcommand{\win}{\textcolor{green!70!black}{$\checkmark$ }}
\newcommand{\winnosig}{\textcolor{orange!90!black}{$\checkmark$ }}
\newcommand{\loss}{\textcolor{red!70!black}{$\times$ }}

\long\def\revision#1{{\color{black}#1}}

\theoremstyle{thmstyleone}
\theoremstyle{thmstyletwo}

\theoremstyle{thmstylethree}

\begin{document}

\title{Separating Ansatz Discovery from Deployment on Larger Problems: Reinforcement Learning for Modular Circuit Design}

\author*[1]{\fnm{Gloria} \sur{Turati}}\email{gloria.turati@polimi.it}

\author[2]{\fnm{Simone} \sur{Foderà}}\email{simone.fodera@lmu.de}
\equalcont{This author contributed while he was a Master's student at Politecnico di Milano.}

\author[1]{\fnm{Riccardo} \sur{Nembrini}}\email{riccardo.nembrini@polimi.it}

\author*[1]{\fnm{Maurizio} \sur{Ferrari Dacrema}}\email{maurizio.ferrari@polimi.it}

\author[1]{\fnm{Paolo} \sur{Cremonesi}}\email{paolo.cremonesi@polimi.it}

\affil[1]{\orgname{Politecnico di Milano}, \city{Milano}, \country{Italy}}
\affil[2]{\orgname{Ludwig Maximilians Universität}, \country{Germany}}


\abstract{
\revision{
As quantum computing continues to gain attention, there is growing interest in how classical machine learning can assist quantum workflows in practice. Automated circuit design, sometimes referred to as Quantum Architecture Search (QAS), is a natural application but relies on the ability to model the quantum system to support learning as the number of qubits grows. This challenge is central to QAS, and much of the current literature that proposes new ways to model the ansatz focuses on small systems, often around ten qubits.
In this work, we propose a complementary approach that separates a small-scale \emph{structure discovery} phase, where a reusable modular circuit block is learned on small instances where classical learning is feasible, from a \emph{deployment} phase, where the blocks are used to create the ansatz required for larger problems.

To this end, we introduce \textit{Reinforcement Learning for Variational Quantum Circuits} (RLVQC), formulating QAS as a sequential decision-making problem. We evaluate our methodology on Quadratic Unconstrained Binary Optimization (QUBO) instances derived from Maximum Cut, Maximum Clique, and Minimum Vertex Cover. 
Our \textit{RLVQC Block} model is trained to discover a modular two-qubit block that can generalize QAOA-style methods and that is often beneficial compared to learning non-modular ansatzes. The blocks discovered on $n=8$ instances remain effective when deployed on larger instances ($n=12$ and $n=16$), supporting the feasibility of reusing learned modular structure across problem sizes.
While we do not aim to establish a new state-of-the-art solver or a computational advantage over classical methods, our results provide evidence that modular ansatz structure can be learned on smaller instances and then extended to larger ones without requiring learning on systems with a large number of qubits, where quantum computing becomes interesting but classical computation becomes impractical.
}
}


\maketitle

\section{Introduction}
\label{sec:introduction}

\revision{
Quantum computing has emerged as a promising way of addressing computational problems that are difficult to solve efficiently on classical hardware. Alongside algorithmic and hardware advances, there is growing interest in how classical machine learning can support quantum computing workflows in practical ways. One relevant direction is the automated design of quantum circuits and ansatzes, sometimes referred to as Quantum Architecture Search (QAS), where data-driven methods can help identify circuit structures that are well matched to a given task. 
However, the fundamental separation between the representational ability of classical learning methods and the exponential growth of quantum systems has so far prevented QAS methods from reaching regimes where quantum computation becomes interesting, with most of the literature focusing on relatively small systems \citep{Zhang_2022,DBLP:conf/icml/WuYLPY23,DBLP:journals/corr/abs-2104-07715,DBLP:conf/iclr/PatelKOBDD24,DBLP:conf/qce/DaiWYC24}. Furthermore, if QAS involves parametric gates, an additional optimization step is required to identify the optimal parameter values tailored to a specific target problem instance. When a new QAS is performed for each problem instance, the computational cost of these two stages may be too great to be justified. Identifying ways to build a machine learning model able to perform QAS on a quantum system of interesting scale is a fundamental research question underpinning QAS that remains widely open. The existing literature has explored several options, with representations based on gate placements~\citep{DBLP:conf/iclr/PatelKOBDD24,DBLP:conf/qce/DaiWYC24,DBLP:journals/natmi/FurrutterMB24}, expectation values~\citep{DBLP:journals/corr/abs-2104-07715}, or even full states and unitaries~\citep{giordano_2022,moro_2021}, each offering advantages and disadvantages. 
In this work, we instead pursue an alternative and complementary approach to address both problems by proposing to split QAS into two phases through modular ansatzes. First, the modular ansatz is learned in a small-scale \emph{structure discovery} phase, at a system size where classical learning is feasible. Then, in the \emph{deployment} phase, the learned modules are used to construct ansatzes suitable for larger problem instances. This strategy both allows to operate in a regime where classical optimization is easier and to separate the two difficult problems of learning the ansatz structure from that of finding the optimal parameters for a specific problem instance. Our focus is on scalable ansatz design and reuse, rather than on establishing a computational advantage over classical solvers.

We evaluate our approach in the context of variational quantum algorithms (VQAs)~\citep{cerezo_2021} in the Noisy Intermediate-Scale Quantum (NISQ) era~\citep{preskill_2018}. While VQAs are not generally associated with provable computational advantages, they provide a widely used test bed in which ansatz design strongly affects solution quality and resource requirements. Indeed, several methods have been proposed to design ansatzes, often by exploiting problem-specific features such as symmetries~\citep{meyer_2023, le_2023, wierichs_2023} or domain structure, \eg in quantum chemistry~\citep{DBLP:conf/icml/WuYLPY23,ostaszewski_2021}. We will focus on QAOA~\citep{farhi_2014} for Quadratic Unconstrained Binary Optimization (QUBO) problems, because it provides a convenient rule to construct ansatzes, from their constituent blocks, for problems of arbitrary size. 

To conduct ansatz discovery, we introduce \textit{Reinforcement Learning for Variational Quantum Circuits} (RLVQC), which models circuit construction as a sequential decision-making process where an agent adds gates based on observations of the current ansatz effectiveness. \textit{RLVQC Block} learns a modular two-qubit block that is used to construct the ansatz in a QAOA-style method, where the blocks are placed according to the relations between the problem variables, yielding a regular and extendable ansatz. We evaluate our method on QUBO instances derived from Maximum Cut, Maximum Clique, and Minimum Vertex Cover, and compare the effectiveness of the modular ansatz with an unconstrained one. This approach opens new research directions to investigate modular block discovery also for tasks beyond combinatorial optimization, such as state preparation.

In summary, our main contributions are as follows:
\begin{enumerate}[label=(\roman*)]
    \item We propose a methodology for scalable ansatz design that separates \emph{structure discovery} from \emph{deployment} by learning a modular circuit block on small instances and reusing it to construct ansatzes for larger instances via an explicit composition rule. This separates two difficult problems in separate stages.
    \item We introduce \textit{RLVQC Block} and assess whether restricting circuit discovery to a modular block structure affects solution quality compared to an unconstrained \textit{Global} variant, showing that this restriction is not detrimental and is often beneficial.
    \item We learn optimal ansatz blocks for problem instances defined on $n=8$ qubits and then test their effectiveness by deploying them to solve larger problems defined on $n=12$ and $n=16$, showing that solution quality remains stable as the problem size increases. The modular ansatzes maintain a favourable trade-off when accounting for the number of parameters and the number of trials required for their optimization, validating the feasibility of our approach. We also assess the effectiveness of the modular ansatzes relative to QAOA and ma-QAOA via statistical significance tests.
\end{enumerate}

The methods, experiments, and results presented in this work extend the findings reported in~\citep{fodera_2024}. The remainder of this paper is organized as follows.
Section~\ref{sec:background} reviews the necessary background and motivates the separation between discovery and deployment.
Section~\ref{sec:rlvqc} introduces our methodology, as well as RLVQC.
Section~\ref{sec:exp_protocol} describes the experimental protocol and benchmarks, and Section~\ref{sec:results} reports the results.
Finally, Section~\ref{sec:conclusions} summarizes the implications of our findings and outlines directions for future work.
}

\section{Background}
\label{sec:background}
\revision{This section introduces the background concepts underlying our approach to modular ansatz design. We first review variational quantum algorithms (VQAs), which provide a standard framework for parameterized circuits and a natural setting for composing repeated circuit blocks according to problem structure (\eg interacting problem variables). We use this setting as a controlled test environment to evaluate whether learned blocks can be reused on larger problem instances. We then review reinforcement learning (RL), which we use to automate the discovery of these circuit blocks, as well as the relevant literature on Quantum Architecture Search.}

\subsection{Variational Quantum Algorithms}
\label{subsec:vqas}

Variational Quantum Algorithms (VQAs)~\citep{cerezo_2021} are a class of hybrid quantum-classical methods designed to solve optimization and simulation tasks. These algorithms employ a parametrized quantum circuit, referred to as the \textit{ansatz}, whose parameters are iteratively adjusted by a classical optimizer to minimize a problem-dependent cost function. The goal is to obtain an optimized ansatz capable of generating the solution to the target problem.
VQAs are particularly well-suited for near-term quantum hardware, as their shallow circuit structure enhances robustness against environmental noise and the effects of limited coherence times. \revision{Nevertheless, VQAs remain heuristic in nature, and there is no known guarantee of a scalable computational advantage over classical optimization algorithms.}

A prominent example of a VQA is the Variational Quantum Eigensolver (VQE)~\citep{peruzzo_2014, tilly_2021}, which has been extensively adopted in quantum chemistry~\citep{cao_2019, mcclean_2016}. VQE estimates the ground state energy of a target Hamiltonian $H$ by preparing a quantum state $\ket{\psi(\theta)}$ through a parametrized circuit and minimizing the expectation value $\langle \psi(\theta) | H | \psi(\theta) \rangle$.
On real quantum hardware, this expectation value is not directly available but must instead be estimated empirically via repeated circuit executions and measurements:
\begin{equation}
    \label{eq:expectation_estimate}
    \langle H \rangle^* = \frac{1}{n_{\text{runs}}} \sum_{i=1}^{n_{\text{runs}}} \langle \tilde{\psi}_i | H | \tilde{\psi}_i \rangle,
\end{equation}
where $n_{\text{runs}}$ denotes the number of runs (\idest executions or shots) of the circuit and $|\tilde{\psi}_i\rangle$ denotes the measured outcome of the $i$-th run.

Another widely studied VQA is the Quantum Approximate Optimization Algorithm (QAOA)~\citep{farhi_2014, blekos_2023}, which has been extensively applied to combinatorial optimization problems~\citep{crooks_2018, willsch_2020, cook_2020, lin_2016, radzihovsky_2019, brandhofer_2022, turati_2022, kurowski_2023}. QAOA constructs its ansatz in layers: it begins with Hadamard gates applied to all qubits, followed by $p$ alternating blocks consisting of a cost unitary operator (derived from the problem Hamiltonian and typically implemented using $RZ$ and $CX$ gates) and a mixer unitary operator that enables a broader exploration of the solution space.

Several variants of QAOA have been proposed to enhance its effectiveness and flexibility. A prominent example is the multi-angle QAOA (ma-QAOA)~\citep{herrman_2022}, which assigns independent parameters to each individual gate, significantly increasing the expressiveness of the variational circuit. However, this gain in flexibility often necessitates a greater allocation of computational resources to navigate the resulting higher-dimensional parameter space.
Another extension, QAOA+, modifies the standard circuit at depth $p=1$ by appending additional problem-independent entangling and mixer layers. By incorporating these extra layers, the algorithm enhances the circuit's capacity to explore the broader solution space effectively.

Despite their potential, VQAs face several critical challenges. A particularly significant obstacle is the emergence of \textit{barren plateaus}~\citep{mcclean_2018, arrasmith_2021, arrasmith_2022, cerezo_2021_bp, holmes_2022, larocca_2022, volkoff_2021}, regions in the parameter space where the gradient of the cost function vanishes exponentially with the number of qubits. This phenomenon limits the effectiveness of optimization methods causing the training to stall and making it difficult to converge to high-quality solutions.
Another major challenge lies in designing effective ansatzes that balance the need for sufficient exploration of the solution space with the requirement of being easily trainable. A well-designed ansatz should exhibit limited circuit depth and gate count, be compatible with the native gate set and connectivity of the target hardware, and possess sufficient expressiveness to capture optimal or near-optimal solutions~\citep{sim_2019, qin_2023, wurtz_2021, du_2020, brozzi_2024}.

To address these challenges, several ansatz design strategies have been proposed. One approach leverages structural properties of the target problem, such as symmetries or interaction topologies~\citep{meyer_2023, le_2023, wierichs_2023, farhi_2014}. Another direction involves \textit{adaptive algorithms}, which build the ansatz incrementally during training by adding or removing gates~\citep{turati_2023, claudino_2020, mukherjee_2023}. These methods start with a smaller and simpler circuit which is then evolved dynamically, avoiding to start with a large and complex one that may be challenging to optimize. 
Adaptive VQAs include an adaptive formulation of VQE, namely ADAPT-VQE~\citep{grimsley_2019}, along with its variants qubit-ADAPT-VQE~\citep{tang_2021}, QEB-ADAPT-VQE~\citep{yordanov_2021}, and Overlap-ADAPT-VQE~\citep{feniou_2023}. These methods are primarily designed for quantum chemistry applications and construct the ansatz by selecting gates from a domain-specific pool tailored to the molecular system under study. 
Adaptive strategies have also been explored for QAOA. For example, the approach proposed in~\citep{zhu_2022} selects mixer unitaries in a layer-wise fashion during training.
Additional adaptive techniques for ansatz construction include genetic algorithms~\citep{rattew_2020, chivilikhin_2020, las-heras_2016}, heuristic optimization methods~\citep{cincio_2018, du_2022, bilkis_2023, ostaszewski_adapt_2021}, and reinforcement learning-based approaches (see Section~\ref{subsubsec:RL_for_QC}), which aim to automate circuit design in a data-driven and systematic way. However, expanding the search space beyond the ansatz parameters to include the structure of the ansatz itself results in a significantly more difficult problem.

\revision{Our work builds on these data-driven approaches to ansatz discovery, while explicitly addressing scalability to larger systems. We propose to separate \emph{structure discovery} from \emph{deployment}: circuit blocks are discovered on small instances where automated search is feasible, and are subsequently composed, via an explicit composition rule, into ansatzes for larger instances. This modular approach enables to build the ansatz required for problems of arbitrary size, avoiding the need to re-run an expensive architecture search whenever the problem dimension increases. More broadly, this research direction targets a key obstacle in automated circuit design, namely the difficulty of scaling architecture search with system size, and seeks to enable classical architecture-search methods for problems of arbitrary (and practically relevant) size.
}

\subsection{Background on Reinforcement Learning}
\label{sec:rl}

This section introduces the core principles of reinforcement learning (RL), with a focus on the specific algorithmic framework adopted in this work.
Throughout the section, we follow standard notational conventions where random variables are denoted by uppercase letters, while their realizations are indicated in lowercase. For a more comprehensive overview of the RL paradigm, the reader is referred to the seminal text by Sutton and Barto~\citep{sutton_1998}.

\subsubsection{States, Actions, and Reward}

\begin{figure}[ht]
  \centering
  \begin{tikzpicture}[
      box/.style={rectangle, draw=black, thick},
      ]
      %Nodes
      \node[box](agent){Agent};
      \node[box](env)[right=of agent]{Environment};
      
      %Lines
      \draw[dashed,->] ([yshift=-0.92cm]$([xshift=5]env.south)!0.8!([xshift=-7]agent.south)$) .. controls +(left:0) and +(down:0.75) .. ([xshift=-7]agent.south)
          node[pos=0.7,left]{$s_t$};
      
      \draw[->] (agent.north) .. controls +(up:1.2) and +(up:1.2) .. (env.north)
          node[pos=0.5,below]{$a_t$};
      
      \draw[->] (env.south) .. controls +(down:1.2) and +(down:1.2) .. (agent.south)
          node[pos=0.5,above]{$s_{t+1}$}
          node[pos=0.5,below]{$r_t$};
  \end{tikzpicture}
  \caption{Interaction between agent and environment in a reinforcement learning framework. At time step $t$ the agent observes the state $s_t$ of the environment, performs an action $a_t$, and receives a reward $r_t$. The environment then transitions to a new state $s_{t+1}$, which the agent observes in the next step. This iterative feedback loop is fundamental to the learning process.}
  \label{fig:agent-env}
\end{figure}

In reinforcement learning (RL), an agent interacts with an environment in a sequence of discrete time steps, indexed by $t$.
At each step, the environment \textit{state} $s_t$ is observed by the agent. Importantly, the agent may not be able to observe the full state $s_t$, therefore states and observations are not necessarily equivalent.
Based on its current observation, the agent selects an \textit{action} $a_t$, which influences the environment and results in a transition to a new state $s_{t+1}$. The agent then receives a scalar \textit{reward} $r_t$, which assesses the quality of the action taken (see Fig.~\ref{fig:agent-env}).
The specific definitions of states, actions, and rewards are problem-dependent and critically affect the agent's capacity to learn how to find high-quality solutions.
The learning process typically consists of multiple \textit{episodes}, where each episode begins from an initial state and proceeds until a predefined termination criterion is satisfied.

The agent's objective is to discover a strategy to maximize the expected cumulative reward, known as the \textit{return}. Formally, the return at time step $t$ is defined as the discounted sum of future rewards:
\begin{equation}
\label{eq:return}
    g_t \coloneq \sum_{k = 0}^{T} \gamma^k r_{t+k+1},
\end{equation}
where $T \in \mathbbm{N} \cup \{+\infty\}$ denotes the number of steps remaining in the episode, and $\gamma \in (0, 1]$ is the \textit{discount factor}. The return is said to have a \textit{finite horizon} if $T \in \mathbbm{N}$, and an \textit{infinite horizon} when $T = \infty$.
The discount factor $\gamma$ modulates the relative importance of immediate versus future rewards. When $\gamma = 1$, all future rewards are equally weighted. As $\gamma$ approaches zero, the agent increasingly emphasizes immediate rewards over long-term gains. Additionally, if $\gamma < 1$ and rewards are bounded, the series in Eq.~\eqref{eq:return} converges even for infinite horizons, owing to the properties of geometric series.

\subsubsection{Policy and Value Function}

In reinforcement learning, the agent's decision-making strategy is encapsulated by a \textit{policy} $\pi$, which defines how actions are selected based on the current state. Policies can be either \textit{deterministic} or \textit{stochastic}.
A deterministic policy maps each state to a single action, producing consistent behavior for the same input:
\begin{equation}
    \pi(s) = a,
\end{equation}
where $s$ is the current state and $a$ is the action determined by the policy.
Conversely, a stochastic policy specifies a probability distribution over actions conditioned on the current state. It is formally defined as:
\begin{equation}
    \pi(a|s) \coloneq P(A_t = a \mid S_t = s),
\end{equation}
where $A_t$ and $S_t$ denote the action and state at time step $t$, respectively.
This formulation allows the agent to exhibit diverse behavior even when revisiting the same state, with action selection governed by the probability distribution $\pi(\cdot|s)$.

To assess the quality of a policy, two fundamental functions are used: the \textit{value function} and the \textit{action-value function}.
The value function $V_\pi(s)$ measures the expected return when starting from state $s$ and subsequently following policy $\pi$:
\begin{equation}
\label{eq:value_function}
    V_\pi(s) \coloneq \mathbb{E}_\pi[G_t \mid S_t = s].
\end{equation}
The action-value function $Q_\pi(s,a)$, on the other hand, estimates the expected return starting from state $s$, taking action $a$, and then following policy $\pi$:
\begin{equation}
\label{eq:action-value_function}
    Q_\pi(s,a) \coloneq \mathbb{E}_\pi[G_t \mid S_t = s, A_t = a].
\end{equation}
These functions provide complementary views of the policy. While $V_\pi(s)$ captures the overall desirability of a state under the current policy, $Q_\pi(s,a)$ offers a more granular perspective by also evaluating the impact of the first individual actions.

\begin{figure}
  \centering
  \begin{tikzpicture}[
          box/.style={minimum width=4mm, minimum height=4mm},
          cell/.style={minimum width=3mm, minimum height=3mm},
          network/.style={minimum width=20mm, minimum height=8mm},
          line/.style={solid},
          dline/.style={densely dashed},
      ]

      \coordinate (base) at (-20mm,-14mm);
      \coordinate (actions) at (30mm,-14mm);
      \coordinate (action) at (42mm,-6mm);
      \coordinate (fillpad) at ($(base) + (4mm,0)$);
      \coordinate (actionpad) at ($(action) + (-4mm,0)$);

      \node[network,draw] (pn) at (0,0) {Policy Net};
      \node[network,draw,below=4mm of pn] (vn) {Value Net};

      \node[fit=(pn)(vn)(fillpad)(actions)(actionpad),inner sep=4mm,draw] (agent) {};
      \node[above=0mm of agent]{Agent};

      \foreach [count=\i from 0,evaluate=\i as \y using \i*4] \shade in {90, 50, 30, 0, 20}{
          \node [box,draw,fill=white] (s\i) at ($(0,\y mm) + (base)$) {};
          \node [cell,draw,fill=magenta!\shade!white,rounded corners=0.8mm] (c\i) at ($(0,\y mm) + (base)$) {};
      }
      \node[box,draw,fill=white,text width=2.7mm,inner sep=0pt] (etc) at ($(0,8mm) + (base)$) {...};
      \draw[white, very thick, dotted] ($(-2mm,6.4mm) + (base)$) -- ($(-2mm,9.6mm) + (base)$) ($(2mm,6.4mm) + (base)$) -- ($(2mm,9.6mm) + (base)$);
      
      \draw[->] ([xshift=-8mm]etc.west) -- (etc.west)
          node[near start,above]{$s_t$};

      \draw[->] (etc.east) .. controls ($(4mm,0) + (etc.east)$) and ($(-4mm,0) + (pn.west)$) .. (pn.west) {};
      \draw[->] (etc.east) .. controls ($(4mm,0) + (etc.east)$) and ($(-4mm,0) + (vn.west)$) .. (vn.west) {};

      \foreach [count=\i from 0,evaluate=\i as \y using \i*4] \shade in {10, 60, 30, 120, 30}{
          \node [box,draw,fill=white] (a\i) at ($(0,\y mm) + (actions)$) {};
          \node [cell,draw,fill=blue!\shade!white,rounded corners=0.8mm] (ca\i) at ($(0,\y mm) + (actions)$) {};
      }
      \node[box,draw,fill=white,text width=2.7mm,inner sep=0pt] at ($(0,8mm) + (actions)$) {...};
      \draw[white, very thick, dotted] ($(-2mm,6.4mm) + (actions)$) -- ($(-2mm,9.6mm) + (actions)$) ($(2mm,6.4mm) + (actions)$) -- ($(2mm,9.6mm) + (actions)$);

      \draw[->] (pn.east) -- ([xshift=-2mm,yshift=14mm]actions.west) {}
          node[below,midway] {$\pi(a|s_t)$};

      \node[right=4mm of vn] (value) {$\hat{V}_\pi(s_t)$};
      \draw[->] (vn.east) -- (value.west) {};
      
      \node[draw,fill=white] (act) at (action) {$a_t$};
      \draw[-] ($(2.8mm,18mm) + (actions)$) .. controls ($(4mm,0) + (a4.north east)$) and ($(-4mm,0) + (act.north west)$) .. ([yshift=-0.07mm]act.north west) {};
      \draw[-] ($(2.8mm,-2mm) + (actions)$) .. controls ($(4mm,0) + (a0.south east)$) and ($(-4mm,0) + (act.south west)$) .. ([yshift=0.07mm]act.south west) {};
      \draw[->] (act.east) -- ([xshift=8mm]act.east);
  \end{tikzpicture}
  \caption{State $s_t$ is processed by the agent's neural networks. The value network outputs an estimate $\hat{V}_\pi(s_t)$ of the value function (\ref{eq:value_function}), while the policy network outputs a probability distribution $\pi(a|s_t)$ on the actions. Action $a_t$ is sampled from this probability distribution.}
  \label{fig:rl_agent}
\end{figure}

\subsubsection{Actor-Critic}
\label{subsubsec:actor-critic}

The actor-critic framework is a reinforcement learning approach based on a dual-network architecture which comprises two main components: the \textit{actor} and the \textit{critic}.
The actor selects the next action to take according to a parametrized policy, which maps observations to actions. It encapsulates the agent's decision-making strategy and guides the agent choice of action in each state.
The critic, on the other hand, evaluates the quality of the actions taken by the actor by estimating either the value function or the action-value function, thereby providing feedback that guides the actor toward more effective policies.

Both components are typically implemented as neural networks, which may either share parameters or not. Each network receives a representation of the environment current state (or observation) as input. The policy network (actor) produces a probability distribution over the action space from which the action is sampled, while the value network (critic) outputs a scalar estimate of the value function at that state (see Fig.~\ref{fig:rl_agent}).

During training, the critic computes the \textit{advantage function}, defined as:
\begin{equation}
    \label{eq:advantage}
    A_\pi(s,a) \coloneq Q_\pi(s,a) - V_\pi(s),
\end{equation}
which measures the relative benefit of taking action $a$ in state $s$, compared to the expected value of the state alone. This advantage estimate is then used to update the parameters of both networks: the critic learns to improve its value estimates, and the actor adjusts the policy to favour actions with higher estimated advantages.

\subsubsection{Proximal Policy Optimization}
\label{subsubsec:ppo}

\textit{Proximal Policy Optimization} (PPO) is a state-of-the-art deep reinforcement learning algorithm introduced by OpenAI~\citep{ppo_2017}.
It is a gradient-based method that employs an actor-critic architecture, where both the policy and value functions are approximated by neural networks (see Fig.~\ref{fig:rl_agent}). PPO optimizes the policy by performing gradient ascent on a surrogate objective, which incorporates an advantage estimate and a regularization mechanism to ensure stable and efficient updates.

One of the key features of PPO is its ability to mitigate the instability commonly associated with traditional policy gradient algorithms. This is achieved through a clipping mechanism in the objective function, which restricts policy updates, thereby enhancing the robustness of the training process.

The clipped surrogate objective optimized by PPO is defined as:
\begin{equation}
    L^{\text{CLIP}}(\theta) \coloneq \hat{\mathbb{E}}_t\left[
    \min\left(
        \rho_t(\theta) \cdot \hat{A}_t, \,
        \text{clip}\left(\rho_t(\theta), 1 - \varepsilon, 1 + \varepsilon\right) \cdot \hat{A}_t
    \right)
    \right],
\end{equation}
where $\hat{A}_t$ denotes an estimate of the advantage function defined in~(\ref{eq:advantage}), and $\rho_t(\theta)$ is the probability ratio between the new and old policies, given by:
\begin{equation}
    \rho_t(\theta) \coloneq \frac{\pi_\theta(a_t \mid s_t)}{\pi_{\theta_{\text{old}}}(a_t \mid s_t)}.
\end{equation}
The ratio $\rho_t(\theta)$ quantifies the difference in the likelihood of taking action $a_t$ under the updated policy $\pi_\theta$ compared to the previous policy $\pi_{\theta_{\text{old}}}$, while the clipping term $\text{clip}(\rho_t(\theta), 1 - \varepsilon, 1 + \varepsilon)$ limits the extent of this change, ensuring that policy updates remain within a trust region.

Thanks to its robustness and adaptability, PPO has become one of the most widely used algorithms in modern reinforcement learning. For further details on PPO's loss formulation and practical implementation, we refer the reader to the original works~\citep{ppo_2017, spinup_2018}.

\subsection{Quantum Architecture Search}
\label{subsubsec:RL_for_QC}

\revision{
Quantum architecture search is not a new task and has been explored for several years with a variety of approaches and objectives. Earlier work includes evolutionary and genetic search methods~\citep{rattew_2020, chivilikhin_2020,las-heras_2016,DBLP:journals/corr/abs-1812-04458,DBLP:journals/corr/abs-2103-06712}. A related line of work includes adaptive and heuristic strategies that build circuits incrementally \citep{cincio_2018, du_2022, bilkis_2023, ostaszewski_adapt_2021,claudino_2020, mukherjee_2023,grimsley_2019,tang_2021,yordanov_2021,feniou_2023}. For example \citep{zhu_2023} proposes ADAPT-QAOA which, via an iterative procedure, selects mixer operators layer by layer, while other similar adaptive construction strategies have been proposed in the context of VQE~\citep{grimsley2019adaptive,PRXQuantum.2.020310}. Among sampling-based methods, researchers have also explored generative machine learning models such as diffusion models~\citep{DBLP:journals/natmi/FurrutterMB24,DBLP:journals/corr/abs-2506-01666}. More recent developments include differentiable quantum architecture search~\citep{Zhang_2022}, which relaxes discrete gate choices into a differentiable probabilistic model and optimizes ansatz selection over a gate pool, with results reported on problems around $n\approx 8$--$10$ qubits. Similarly, QuantumDARTS~\citep{DBLP:conf/icml/WuYLPY23} uses a Gumbel--Softmax relaxation to sample gates in a differentiable way and proposes learning small sub-circuits that are then stacked to form the full circuit, evaluating on quantum chemistry and optimization problems up to $n=10$ qubits.

In recent years, reinforcement learning (RL) has been increasingly applied in quantum computing, addressing tasks where the search space is large and the objective can be evaluated through interaction with a simulator or hardware. A first area of interest is the optimization of quantum circuit parameters~\citep{wauters_2020, khairy_2020, yao_2020}. RL has also been used for quantum circuit compilation and related preprocessing tasks, where the goal is to transform a circuit into a form that better matches hardware constraints. This includes reducing circuit depth and gate count~\citep{fosel_2021}, as well as qubit routing~\citep{DBLP:conf/aaai/SinhaAS22, DBLP:journals/tqc/PozziHSM22}, where the goal is to modify a circuit to match the connectivity of the target device, and compilation methods based on sequential decision making~\citep{zhang_2020, moro_2021}.

For the purposes of this work, we focus on RL applied to quantum architecture search. Among early examples, \citep{DBLP:journals/corr/abs-2104-07715} trains an RL agent for GHZ state preparation on 2- and 3-qubit systems under both ideal and noisy conditions. The agent observes single-qubit Pauli expectation values, yielding an observation vector that scales linearly as $3n$ with the number of qubits. More broadly, RL-based circuit synthesis, where an agent generates a circuit that maps an initial state to a target state, has been explored in a range of settings. For example, \citep{giordano_2022} employs Q-learning to generate four-qubit entangled states, \citep{gqpr-dgz7} applies RL to learn fault-tolerant circuits for quantum error correction, and \citep{11032620} focuses on the synthesis of Clifford operators. RL has also been applied to the automated design of parametrized circuits for classification tasks~\citep{pirhooshyaran_2021}.

Another line of work focuses on ansatz search for VQE-style problems. \citep{DBLP:conf/iclr/PatelKOBDD24} proposes a curriculum RL approach for quantum architecture search in VQE, targeting molecular ground-state energy estimation up to 8 qubits, and represents the environment state via a tensor-based (3D) binary encoding of the circuit architecture, building on \citep{ostaszewski_2021}. In a similar direction, \citep{DBLP:conf/qce/DaiWYC24} uses an RL agent with a circuit-encoding observation represented as a flattened $4 \times L$ tensor, where $L$ denotes the maximum circuit depth, and applies it to ansatz discovery for quantum classification tasks, with experiments up to 4 qubits. RL has also been combined with QAOA-style heuristics. In particular, \citep{DBLP:journals/corr/abs-2207-06294} proposes an RL-enhanced variant of recursive QAOA (RQAOA) that learns decisions in the recursive variable-elimination procedure. This improves the overall heuristic, but differs from our focus since it does not aim to discover a reusable modular circuit block that can be composed into an ansatz for larger instances.

A common pattern across current QAS and RL-based circuit synthesis methods is that empirical evaluations are typically limited to very small systems, often up to $n\approx 8$ qubits and, in some cases, around $n\approx 10$. This suggests that scalability remains a central challenge and that many of these approaches have yet to demonstrate suitability in regimes that are classically challenging and therefore of greatest practical interest. A key obstacle is the definition of the agent observation due to how a faithful description of a quantum state scales exponentially with the number of qubits. Existing works therefore rely on task-specific encodings, including explicit circuit encodings such as binary tensor representations of gate placements~\citep{DBLP:conf/iclr/PatelKOBDD24,DBLP:conf/qce/DaiWYC24,DBLP:journals/natmi/FurrutterMB24}, summaries based on Pauli expectation values~\citep{DBLP:journals/corr/abs-2104-07715}, or, in some cases, full state~\citep{giordano_2022} or unitary~\citep{moro_2021} representations. Each option introduces different trade-offs. Circuit encodings based on gate positions avoid exponential scaling but are not unique, since the same circuit can admit different representations depending how tightly the gates have been parallelized, and they do not natively capture equivalences between gate sequences. As a result, the learning algorithm must implicitly discover these relations from data, which can make training brittle and generalization to new ansatzes difficult, especially as the number of qubits increases. Observations based on expectation values can scale more favourably, are naturally compatible with sampling on real hardware, and avoid non-unique circuit encodings, but they may provide too limited a view of the computation to remain effective as system size grows, potentially causing learning to stall. Using full state or full measurement statistics can be informative at small $n$ but becomes rapidly intractable as $n$ increases. Identifying observation representations that remain informative while scaling favourably up to a number of qubits where quantum computation becomes interesting is therefore an open and essential research question underpinning the entire QAS task, and the answer will likely depend on the system size and the scenario of interest.

A further practical consideration is that end-to-end QAS compounds two difficult optimization problems when applied to tasks that require parametric gates: selecting the gate parameters and selecting the circuit architecture. On small instances, architecture search methods that require many interaction steps, together with repeated gate parameter optimization, can lead to an overall number of circuit evaluations that is difficult to justify, and may even exceed the cost of solving the problem directly (\eg via enumeration). This motivates our methodology, which separates a small-scale \emph{structure discovery} phase from a \emph{deployment} phase. We use RL only where learning is feasible to discover modular ansatzes in the form of reusable circuit blocks, and then instantiate these blocks via an explicit composition rule to construct ansatzes for larger instances without requiring the agent to operate in that large-dimensional regime. This split has two practical advantages. First, the discovery phase can use as much computation and optimization as needed, since it is decoupled from the large instances we ultimately care about. Second, learning on a small system avoids the worst scalability issues and lets us use simpler observations that tend to behave more predictably.

To the best of our knowledge, this work is the first to investigate an automated synthesis method that identifies an ansatz block on a reduced number of qubits and then extends the composed ansatz to solve larger problem instances. Learning a modular structure that can be instantiated for problems of increasing size represents a concrete step toward addressing the scalability limitations inherent in automated variational circuit design.

}

\subsection{QUBO Problems}
\label{subsec:qubo}

Quadratic Unconstrained Binary Optimization (QUBO) problems are NP-hard combinatorial optimization tasks, where the goal is to find a binary vector that minimizes a quadratic cost function. Formally, a QUBO problem is defined as:
\begin{equation}
    \min_{x \in \{0,1\}^n} x^\top Q x,
\end{equation}
where $x \in \{0,1\}^n$ is a binary vector and $Q \in \mathbb{R}^{n \times n}$ is a symmetric (or upper triangular) matrix that encodes the cost landscape.

The QUBO formulation allows to represent rather easily many combinatorial optimization problems, often defined on graphs, such as Maximum Cut, Minimum Vertex Cover, and Maximum Clique \citep{Glover2022,lucas2014ising}. These problems are described in more detail in Appendix~\ref{app:problems_description}. Although QUBO problems are, by definition, unconstrained, many real-world formulations involve constraints. This presents a challenge when converting such problems into QUBO form. A common strategy to overcome this issue is to incorporate the constraints as penalty terms in the objective function, increasing the cost for any solution that violates them.

QUBO problems can also be reformulated as Ising models by mapping binary variables to spin variables. This transformation enables the use of quantum algorithms such as VQE and QAOA, which are specifically designed to approximate the ground state of the corresponding Hamiltonian, but also other special purpose quantum hardware that is developed specifically to minimize Hamiltonians of this type. Note that QUBO formulations have been applied in many fields for applied problems, such as machine learning \citep{Neukart2018, Mott2017, Mandra2016, DBLP:conf/qce/CarugnoDC24, DBLP:conf/sigir/DacremaMN0FC22, Neven2009, WillschWRM20, KumarBTD18, Neukart2018_cluster, OMalleyVAA17, ottaviani2018low, DBLP:conf/recsys/NembriniCDC22, nembrini_cqfs}, chemistry \citep{Micheletti2021, HernandezA17, streif2019solving, Xia2018}, optimization and logistics \citep{Ikeda2019, RieffelVODPS15, Ohzeki2020, Carugno2022, StollenwerkLJ17,DBLP:conf/qce/ChiavassaMPDC22}, highlighting the importance of a simple and flexible formulation that can accommodate many types of tasks and is suitable for different quantum computing platforms.

\revision{In addition to its broad applicability, the QUBO form provides a convenient \emph{composition rule} for modular ansatzes similar to what is used in QAOA: the quadratic couplings define an interaction structure over variables (equivalently, a weighted graph), where each non-zero term $q_{ij}$ identifies an interacting pair. In our setting, this interaction structure is used to instantiate and repeat learned circuit blocks across all interacting qubit pairs, enabling a direct way to scale a learned block from small instances to larger ones.
}

\section{RLVQC Framework: Structure Discovery and Deployment}
\label{sec:rlvqc}
\revision{The overall goal of this study is to explore the feasibility of learning modular ansatzes via classical machine learning methods in regimes where this is possible, and then reusing the discovered blocks to tackle larger problems. We therefore aim to separate the phase of \emph{discovery} from that of \emph{deployment}. This separation has the potential to leverage classical learning to optimize ansatz design also for qubit sizes that are impractical to simulate, or beyond classical simulation altogether. To achieve this, we require two components: (i) a learning method to optimize the ansatz and discover effective blocks, and (ii) a composition rule to define the ansatz for a larger problem based on the learned blocks.
As previously stated, we focus on QUBO optimization with QAOA-style methods because they offer a convenient way to compose the ansatz and are a widely used test bed. To learn the ansatz, we introduce our reinforcement learning-based algorithm, \textit{Reinforcement Learning for Variational Quantum Circuits} (RLVQC), presented in two variants: \textit{Block} and \textit{Global}. The algorithm aims to construct quantum circuits that approximate the ground state of a Hamiltonian corresponding to a given QUBO instance. The task is formulated as a sequential decision-making problem, where the RL agent iteratively adds gates to the ansatz according to the policy it has learned. We provide an overview of the key components of RLVQC, including the environment, action space, and reward function, followed by a description of the agent's training procedure. Finally, we present three parameter-sharing variants for the blocks (\texttt{Agnostic}, \texttt{Weighted}, and \texttt{Tied}), and describe how the discovered blocks are scaled and instantiated to solve larger problem instances.}

\subsection{Key Components}
\label{sec:rlvqc:components}

The methods we propose, RLVQC Block and Global, are based on the same underlying architecture but differ in the definition of the action space. In the following, we provide a detailed description of each component.

\paragraph{Environment}
The environment consists of a parametrized quantum circuit acting on $n$ qubits. The initial state corresponds to a circuit where a single layer of Hadamard gates is applied to all qubits. 
\revision{At each step, the agent selects a gate which is appended to the circuit, thereby increasing its depth. As actions accumulate, the circuit grows longer and contains an increasing number of gates. Note that the actual ansatz that is run will depend on the experiment: as we will explain later, while RLVQC \textit{Global} learns the actual ansatz, \textit{Block} constructs a 2-qubit block that will be used to build the ansatz.}

\paragraph{Observations}  
\revision{When defining observations for this task, we can take several routes, as discussed in Section \ref{subsubsec:RL_for_QC}, \eg tensor representations of gate placements~\citep{DBLP:conf/iclr/PatelKOBDD24,DBLP:conf/qce/DaiWYC24,DBLP:journals/natmi/FurrutterMB24}, expectation values~\citep{DBLP:journals/corr/abs-2104-07715}, or full state or unitary representations~\citep{giordano_2022,moro_2021}. Given that determining the most appropriate observation for a given system is still an open research question and that, as we have highlighted earlier, representations based on expectations or the ansatz itself could introduce confounding factors, we derive our observations from the full state, in line with our goal of testing the feasibility of learning ansatz blocks during the \emph{discovery} phase on systems that can be simulated. Still, the development of an effective model for the quantum circuit to allow QAS on larger systems is a valuable complementary direction that would benefit our proposed methodology as well.

}
 
We define an observation as a $2^n$-dimensional vector containing empirical estimates of the probabilities of measuring each computational basis state. Given a quantum state $\ket{\psi} = \sum_i c_i \ket{i}$, where $c_i \in \mathbb{C}$ and $\ket{i}$ denotes a computational basis state, the probability of observing outcome $\ket{i}$ upon measurement is $\abs{c_i}^2$. Since the exact amplitudes $c_i$ are not directly accessible, we estimate these probabilities by executing the circuit $n_{\text{runs}}$ times and measuring the resulting outputs. Let $n_i$ be the number of times the state $\ket{i}$ is observed. The corresponding empirical frequency $\hat{p}_i = n_i / n_{\text{runs}}$ provides an estimate of $\abs{c_i}^2$. Thus, the complete observation vector passed to the agent is given by $[\hat{p}_0, \dots, \hat{p}_{2^n-1}]$.

This scenario is an example of partially observable environment state. While it would be possible to use as observations the complex amplitude values computed during the simulation, we aim to keep our work closer to a realistic scenario where the environment could be a real quantum computer and, as such, the amplitude values would not be directly available. Note that it would also be possible to estimate the amplitudes by other means, such as using state tomography, but this would also introduce additional complexity and its own set of approximations.

\paragraph{Reward}

The reward function is designed to simultaneously minimize the Hamiltonian expectation value and the circuit depth. It is formally defined as: 
\begin{equation}
\label{eq:reward}
    R_t \coloneq -\langle H\rangle^*_t - \beta d_t,
\end{equation}
where $\langle H\rangle_{t}^{*}$ is the estimated expectation value at time $t$ (as defined in~(\ref{eq:expectation_estimate})), and $d_{t}$ denotes the current circuit depth, calculated with respect to a specified basis gate set. The parameter $\beta$ is a penalty coefficient that controls the trade-off between energy minimization and circuit shallowness. 
This trade-off can be chosen depending on the specific scenario of interest.

The expectation value for the circuit is estimated empirically by executing the circuit $n_{\text{runs}}$ times and averaging the energies of the resulting measurement outcomes. This empirical approach is chosen to reflect the behaviour of real quantum hardware, where exact computation of the expectation value is typically infeasible.

This reward formulation incentivizes the agent to construct ansatzes that are both effective and hardware-efficient.
This definition of the reward function is intentionally simple and directly aligned with the core task objectives.
Nonetheless, reward design in reinforcement learning is highly flexible, and alternative reward functions could be adopted to prioritize different algorithmic or hardware constraints.

\paragraph{Actions}
\revision{
Our main goal is to learn modular ansatzes whose blocks can be used to instantiate ansatzes for larger problems. This is achieved by \textit{RLVQC Block}, which constructs a modular two-qubit block that is subsequently composed to build the final ansatz. However, since this introduces an explicit structural restriction and the \textit{Block} variant operates in a substantially more constrained search space, we also consider a reference variant that constructs the complete circuit directly, without predefined structural constraints, namely \textit{RLVQC Global}. Comparing these two variants allows us to assess the impact of imposing modularity on the learned ansatzes.

In both models, an action is defined as the addition of a single gate to either the current two-qubit block (in RLVQC Block) or the current circuit (in RLVQC Global).
Each action specifies both the gate type and the target qubit(s). Consequently, the action space $\mathcal{A}$ comprises all valid combinations of gates and qubits available to the agent within its respective scope. The gates for both variants may include both single-qubit and two-qubit gates, which can be either parametric or non-parametric. 
Note that the specific gate set from which the agent can choose depends on the experiment and will be explained in Section \ref{sec:exp_protocol:effectiveness_block} and \ref{sec:exp_protocol:block_transfer}.
}

The main distinction between the two variants lies in the configuration of this action space:
\begin{itemize}
    \item \textbf{RLVQC Block}: This variant enforces an ansatz structure inspired by QAOA. An action corresponds to inserting a gate into a foundational two-qubit block. At each step, the current block is applied to all \textit{interacting qubit pairs}\footnote{Interacting qubits are defined as pairs $(i, j)$ corresponding to binary variables with non-zero interactions in the QUBO cost function ($q_{ij} \neq 0$).} to obtain a candidate $n$-qubit ansatz. The parameters of the gates may be independent or not depending on the specific parameter-sharing scheme, which is described in Section~\ref{sec:rlvqc:param_sharing}.
    \item \textbf{RLVQC Global}: This variant removes all architectural constraints, allowing the agent to place gates between any available qubit pair or on any individual qubit. This offers maximal flexibility in circuit design at the cost of a larger search space. Here, an action is defined as the insertion of a specific gate applied to one or two selected qubits among the total number of qubits available.
\end{itemize}

\paragraph{Agent Architecture}

RLVQC employs the PPO framework described in Section~\ref{subsubsec:ppo}, which is based on two neural networks: a policy network and a value network. Both networks are implemented as fully connected, multi-layer feed-forward networks with identical architectures, except for the output layer. The networks maintain independent sets of learnable parameters. 

The input layer of both networks consists of $2^n$ neurons, where $n$ is the number of qubits. This corresponds to the dimensionality of the observation vector provided to the agent (see Section~\ref{sec:rlvqc:components}). \revision{Again, we stress that other architectures could be used by representing the environment observation in a different way. However, this would cause the introduction of other approximations and in this work we aimed at reducing the number of confounding factors. Furthermore, even if the agent architecture would not scale to large system, this is consistent with our goal of separating the phase of ansatz discovery on small-to-medium problems, from that of deployment onto large ones.}

The size of the output layer instead depends on the role of each network. For the value network, the output is a single scalar representing the estimated value of the current state. For the policy network, the output layer size equals the cardinality of the action set, \idest $\abs{\mathcal{A}}$, which varies depending on the algorithm variant. In RLVQC Global, $\abs{\mathcal{A}}$ depends on both the number of qubits and the available gates. In contrast, for RLVQC Block, it depends only on the number of gates, as the qubit indices are two by construction.

\begin{figure}
  \centering
        \begin{tikzpicture}[
                box/.style={minimum width=4mm, minimum height=4mm},
                cell/.style={minimum width=3mm, minimum height=3mm},
                network/.style={minimum width=20mm, minimum height=8mm},
                line/.style={solid},
                dline/.style={densely dashed},
            ]
            \node[scale=0.8] (circ) {
            \begin{quantikz}[row sep=0.2em, column sep=0.8em]
                \lstick{$q_2 :$} & \gate{\mathrm{H}} & \ \ldots\ & \ctrl{1} & \qw & \ctrl{1} & \qw & \meter{} \\
                % ..........
                \lstick{$q_1 :$} & \gate{\mathrm{H}} & \ \ldots\ & \targ{} & \gate{R_z(\theta_{t-1})} & \targ{} & \qw & \meter{} \\
                % ..........
                \lstick{$q_0 :$} & \gate{\mathrm{H}} & \ \ldots\ & \qw & \qw & \qw & \qw & \meter{}
            \end{quantikz}
            };
            \node[above left=-2mm of circ,anchor=south west] (ts) {Time step $t-1$};
            \node[above=0mm of ts] (fillts) {};
    
            \node[scale=0.8,below=6mm of circ] (postcirc) {
            \begin{quantikz}[row sep=0.2em, column sep=0.8em]
                \lstick{$q_2 :$} & \gate{\mathrm{H}} & \ \ldots\ & \ctrl{1} & \qw & \ctrl{1} & \qw & \qw & \qw & \meter{} \\
                % ..........
                \lstick{$q_1 :$} & \gate{\mathrm{H}} & \ \ldots\ & \targ{} & \gate{R_z(\theta_{t-1})} & \targ{} & \qw & \qw & \qw & \meter{} \\
                % ..........
                \lstick{$q_0 :$} & \gate{\mathrm{H}} & \ \ldots\ & \qw & \qw & \qw & \qw & \gate{R_x(\theta_t)} \gategroup[1,steps=1,style={dashed},label style={label position=below,yshift=-4mm}]{$a_t$} & \qw & \meter{}
            \end{quantikz}
            };
            \node[above left=-2mm of postcirc,anchor=south west]{Time step $t$};
    
            \draw[->] (circ.south) -- ([yshift=-2mm]postcirc.north) {};
    
            \node[box,inner sep=2mm,below=4mm of postcirc,align=center,draw] (opt) {Optimization\\+\\Simulation};
            \draw[->] ([yshift=6mm]postcirc.south) -- ([yshift=2mm]opt.north) {};
    
            \node[fit=(fillts)(circ)(postcirc)(opt),inner sep=4mm,draw] (env) {};
            \node[above=0mm of env]{Environment};
    
            \node[box,draw,fill=white,above left=2mm of fillts] (action) at (fillts) {$R_x(\theta_t)$};
            \draw[->] ([yshift=8mm]action.north) -- (action.north) {}
                node[left,midway] {$a_t$};
    
            \node[draw,box,below right=0mm and 8mm of opt,inner sep=2mm,outer sep=2mm,fill=white,align=center] (state) {$s_{t+1}$\\$r_t$};
            \draw[->] ([xshift=2mm]opt.east) .. controls ($(opt) + (27mm,0)$) and ($(state) + (0,15mm)$) .. (state.north) {};
            \draw[->] ([xshift=-2mm]state.east) -- ([xshift=8mm]state.east) {};
    
        \end{tikzpicture}
        \caption{When the environment receives action $a_t$, the corresponding gate is added to the circuit. Then, its parameters are optimized and the circuit is simulated to obtain the next state $s_{t+1}$, which is sent back to the agent with the corresponding reward $r_t$.}
        \label{fig:rl_env}
    \end{figure}

\subsection{Parameter-Sharing Variants}
\label{sec:rlvqc:param_sharing}
\revision{
In this section, we define three parameter-sharing schemes for RLVQC Block. These variants are distinguished by their parameterization strategies and the inclusion of problem-specific scaling coefficients:
\begin{itemize}
    \item \textbf{Agnostic}: In this configuration, every gate within the ansatz is independently parameterized. This approach provides maximum expressivity, similar to multi-angle QAOA (ma-QAOA).
    \item \textbf{Weighted}: Each gate instance remains independently parameterized, but the rotation angle $\theta$ is scaled by the corresponding interaction coefficient $q_{ij}$ derived from the QUBO problem. This variant aims to embed local problem information directly into the ansatz structure.
    \item \textbf{Tied}: This variant introduces parameter tying across the circuit. While each rotation angle $\theta$ is scaled by the interaction coefficient $q_{ij}$, the underlying parameter is shared across all blocks within the same layer. For example, if the block contains two rotation gates, there will only be two free parameters per layer, one for the first gate and one for the second, regardless for how many times the block is replicated in the final ansatz. This parameter-sharing scheme is similar to QAOA and significantly reduces the total number of parameters to optimize, thereby reducing computational resources.
\end{itemize}
The objective of these variants is twofold: first, to evaluate whether incorporating structural information from the Hamiltonian improves the agent's convergence; and second, to determine if the reduction in parameter dimensionality achieved through parameter tying can maintain high-quality solutions while improving training efficiency.
}

\subsection{Episodes}
\label{sec:rlvqc:episodes}

Within the RLVQC framework, an episode consists in the iterative construction of a single ansatz. Each episode begins by initializing the quantum circuit with a layer of Hadamard gates applied to all qubits, establishing an initial state of uniform superposition. The episode then proceeds through a series of discrete time action steps, each characterized by a nested two-stage optimization process.

In the primary stage, the RL agent performs its sequential decision-making task. At each time step $t$, the agent selects a gate from the action space and appends it to the existing circuit. For parametric gates, the operation is initialized to default initial value. 

The secondary stage consists of a classical optimization routine. Here, the parameters of the gates currently in the circuit are refined to minimize the cost function, which is defined as the empirical expectation value of the problem-specific Hamiltonian (see~(\ref{eq:expectation_estimate})). Note that we use different parameter-sharing schemes (see Section \ref{sec:rlvqc:param_sharing}).
Following this local optimization, the circuit is executed to obtain measurement outcomes, used to estimate the state probabilities which are used to form the observation vector for the subsequent step $t+1$.

This iterative loop continues until a predefined termination criterion is satisfied, such as the agent reaching a maximum number of steps or the reward exhibiting a plateau in improvement. \revision{The details of the termination conditions depend on the experiments and are reported in Section \ref{sec:exp_protocol:effectiveness_block} and \ref{sec:exp_protocol:block_transfer}.}
Upon the conclusion of an episode, the circuit is reset to its initial state, and a new episode starts to explore alternative ansatz structures
A conceptual overview of this process is illustrated in Fig.~\ref{fig:rl_env}.

\subsection{Agent Training and Final Ansatz}
\label{sec:rlvqc:training_and_final}

The agent training process is organized as a series of episodes, each consisting in the sequential construction of an ansatz through successive action steps, as we previously described. As the episodes progress, the parameters of the agent's policy and value networks are periodically updated every fixed number of steps, called an \textit{epoch}, whose length is a hyperparameter.

The training concludes once a termination condition is reached, which is typically defined by a predetermined budget of interaction steps. 
At the end of the training, we select the circuit that achieves the highest reward among all the explored ones, rather than the final circuit from the last training step. \revision{This concludes the \emph{discovery} phase.}

\revision{
When using \emph{RLVQC Block} to learn a block that will be used in the \emph{deployment} phase for larger problems (see Section \ref{sec:exp_protocol:block_transfer}), we then need to compose the final ansatz which will be used to run the experiment. The composition works as follows. The ansatz starts with a layer of Hadamard gates applied to all qubits. Then, the ansatz is built in $p$ layers composed first by applying the optimal block discovered by the agent to all interacting qubit pairs, which are defined by the interaction terms in the QUBO problem. Then, if $p > 1$, successive layers are separated by a layer of $R_X$ gates, following the structural logic of QAOA. Note that $p$ is a hyperparameter. The gates are parametrized following the parameter-sharing scheme the block was learned with, see Section \ref{sec:rlvqc:param_sharing}.
}

\section{Experimental Protocol}
\label{sec:exp_protocol}

\revision{
This section details the experimental protocol used to train the RLVQC agents and evaluate their solution quality. Our overarching goal is to assess a two-phase methodology that separates \emph{structure discovery} from \emph{deployment}: we use the RLVQC agent to discover modular ansatz blocks on small instances, and then use these blocks to construct ansatzes for larger instances where direct ansatz structure search is not intended to be performed.
Our evaluation is therefore organized around two primary objectives:
\begin{itemize}
    \item \textbf{Experiment 1: Effectiveness of the Block Structure}. We first assess whether introducing an explicit block structure, and the associated reduction in action space, negatively impacts learning. We compare \textit{RLVQC Block} against \textit{RLVQC Global} and standard QAOA in terms of solution quality and circuit efficiency, evaluating whether the modular restriction is neutral or beneficial.
    \item \textbf{Experiment 2: Extending the Modular Ansatz to Larger Instances}. We then study the deployment phase, investigating whether blocks discovered on small-scale instances remain effective when instantiated on larger, more complex instances via the prescribed composition rule. In this setting, we evaluate the three \textit{RLVQC Block} parameter-sharing variants and compare against standard QAOA and ma-QAOA baselines.
\end{itemize}
Both RLVQC variants and all baselines are evaluated on a consistent set of benchmark problem instances.
}

The remainder of this section is organized as follows. First, we characterize the problem instances and the graph topologies considered in our evaluation. Subsequently, we introduce the evaluation metrics utilized to quantify the effectiveness of the proposed methods. Finally, we provide specific implementation details for both experiments, specifying the gate sets, termination conditions, and the computational budget allocated for agent training and parameter optimization.

\subsection{Problem Instances}
\label{sec:exp_protocol:instances}

Our experiments are designed to evaluate the effectiveness of RLVQC in solving optimization tasks formulated as Quadratic Unconstrained Binary Optimization (QUBO) problems (see Section~\ref{subsec:qubo}). 
Recall that QUBO problems can be reformulated as Ising models through a suitable change of variables. In this formulation, the objective becomes identifying the ground state of a Hamiltonian operator, a task that aligns well with the capabilities of both RLVQC and QAOA.

We consider three representative combinatorial optimization problems: Maximum Cut, Maximum Clique, and Minimum Vertex Cover. Each of these can be formulated in QUBO form~\citep{DBLP:journals/anor/GloverKHD22, pelofske_2019}, see Appendix~\ref{app:problems_description}. The specific problem instances are derived from graphs of varying sizes, specifically with 8, 12, and 16 vertices, and span a diverse set of topologies. Note that given the type of problems, the number of vertices $n$ coincides with the number of qubits required.
The eight graph topologies used in our experiments include ``3-regular'', ``Erdős–Rényi'' with edge probabilities 0.2 and 0.7, and ``Barabási–Albert'' with parameters $m = 0.2\,n$ and $m = 0.5\,n$ (where $n$ is the number of vertices and $m$ is the number of edges each new node creates). We also use ``2d-grid'' topologies with $m = 4$ vertices per side\footnote{Note that $m = 4$ evenly divides all selected sizes $n = 8,\ 12,\ 16$, yielding 2d-grid topologies of dimensions $4\times2$, $4\times3$, and $4\times4$, respectively.}, as well as ``star'' and ``cycle'' graphs.

For each combination of topology and graph size, we generate a new graph\footnote{All graphs are constructed using the \texttt{NetworkX} Python library available at~\url{https://networkx.org/}. We ensure the graphs are connected. For random topologies that depend on a random seed, we use the smallest seed value that yields a connected graph, to ensure reproducibility.}, resulting in a total of 24 graphs. From each graph, we compute the QUBO formulation of each of the three optimization problems, yielding 72 QUBO instances in total.

It is worth noting that both the Maximum Clique and Minimum Vertex Cover problems involve constraints, which are incorporated into the QUBO formulation via penalty terms added to the objective function. Each penalty term evaluates to 1 when a constraint is violated and 0 otherwise, and is scaled by a suitable weight coefficient. The penalty weights are chosen so that any feasible solution always attains a lower cost than any infeasible one, thereby guiding the algorithm toward the selection of feasible solutions.

\subsection{Evaluation Metrics}
\label{sec:exp_protocol:metrics}

The primary objective of this study is to assess whether an RL agent can design ansatzes capable of sampling high-quality solutions to optimization problems. To evaluate its effectiveness, we consider both the quality of the solutions produced and the structural characteristics of the resulting circuits.

\paragraph{Approximation Ratio}
The primary metric used to evaluate solution quality is a normalized version of the \textit{approximation ratio} (A.R.), which accounts for shot noise in the expectation estimate and includes a normalization step. The normalized approximation ratio is defined as:
\begin{equation}
    \label{eq:normalized_approx_ratio}
    \text{A.R.} \coloneq \frac{\langle H \rangle^* - \langle H \rangle_{\text{max}}}{\langle H \rangle_{\text{min}} - \langle H \rangle_{\text{max}}},
\end{equation}
where $\langle H \rangle_{\text{min}}$ and $\langle H \rangle_{\text{max}}$ denote the minimum and maximum attainable expectation values of the cost Hamiltonian, respectively, and $\langle H \rangle^*$ is the estimated expectation value, as defined in~(\ref{eq:expectation_estimate}).

This metric approaches 1 when $\langle H \rangle^*$ is close to $\langle H \rangle_{\text{min}}$, indicating that the circuit effectively samples states with near-optimal energy. To ensure that the ratio remains confined to the interval $[0, 1]$, the normalization step subtracts $\langle H \rangle_{\text{max}}$ from both the numerator and the denominator. 
It is worth noting that, for the Maximum Cut problem, $\langle H \rangle_{\text{max}} = 0$.

Finally, it is important to emphasize that the normalized approximation ratio requires prior knowledge of both the minimum and maximum expectation values. Therefore, this metric is applicable only for benchmarking purposes.

\paragraph{Circuit Composition}
To further characterize the learned circuits, we examine the composition of the final circuits, focusing on metrics such as total gate count and overall circuit depth. Note that the two may be different as multiple gates may be applied in parallel. Additionally, we report the number of gates of each type. 
These metrics are computed using the circuits directly as generated by the algorithms, without any modifications, such as simplification or changes in the gate basis. 

\subsection{Experiment 1: Effectiveness of the Block Structure}
\label{sec:exp_protocol:effectiveness_block}

\revision{
This experiment assesses the effectiveness of the modular block structure, compared with both the unconstrained \textit{RLVQC Global} variant and the standard QAOA baseline, in terms of solution quality and circuit resources. Since our broader goal is to enable a separation between a small-scale \emph{structure discovery} phase and a large-scale \emph{deployment} phase, we first need to verify that introducing modularity does not hinder learning. In particular, \textit{RLVQC Block} operates in a substantially more constrained action space, as it learns a relatively small two-qubit block that is then composed to form the full ansatz. This experiment therefore evaluates whether this restriction is detrimental relative to learning an unconstrained circuit directly. We conduct this experiment using the \texttt{Agnostic} parameter-sharing scheme, to facilitate a more direct comparison with \textit{RLVQC Global}. In order to assess the variance we repeat each experiment 5 times.
}

\subsubsection{Gate Set}
The gate set used in the action space of this first experiment includes both single-qubit and two-qubit gates, along with the indices of the qubits on which they act. 
Therefore, the action set is given by $\mathcal{A} = \mathcal{S} \cup \mathcal{D}$, where the sets $\mathcal{S}$ and $\mathcal{D}$ are defined as follows:
\begin{itemize}
    \item \textbf{Single-qubit rotation gates ($\mathcal{S}$):}
    \begin{equation}
    \label{eq:single_rotations}
        \mathcal{S} \coloneq \{R_a^i(\theta) \mid a \in \{x, y, z\}, \ i = 0, \dots, n - 1\},
    \end{equation}
    where $R_a^i(\theta)$ denotes a rotation by angle $\theta$ around the $a$-axis applied to qubit $i$.

    \item \textbf{Two-qubit rotation gates ($\mathcal{D}$):}
    \begin{equation}
    \label{eq:double_rotations}
        \mathcal{D} \coloneq \{R_{ab}^{ij}(\theta) \mid a, b \in \{x, y, z\}, \ i, j = 0, \dots, n - 1, \ i < j\},
    \end{equation}
    where $R_{ab}^{ij}(\theta)$ applies the two-qubit operator $R_{ab}(\theta)$ to qubits $i$ and $j$ and can be used to introduce entanglement.
 The two-qubit rotation gate $R_{ab}(\theta)$ is defined as:
\begin{equation}
\label{eq:double_rotation_definition}
    R_{ab}(\theta) = e^{-i \frac{\theta}{2} \, \sigma_a \otimes \sigma_b},
\end{equation}
where $\sigma_a$ and $\sigma_b$ are Pauli matrices along axes $a$ and $b$, respectively, with $a, b \in \{x, y, z\}$.
This operator can be implemented through the following decomposition:
\begin{equation}
\label{eq:double_rotations_decomposition}
    R_{ab}(\theta) = (U_a^\dagger \otimes U_b^\dagger) \, R_{zz}(\theta) \, (U_a \otimes U_b),
\end{equation}
\begin{equation*}
    \vcenter{\hbox{
        \begin{quantikz}
        & \gate[2]{R_{ab}(\theta)} & \qw \\
        & \qw & \qw
        \end{quantikz}
    }}
    \ = \
    \vcenter{\hbox{
        \begin{quantikz}
        & \gate{U_b} & \gate[2]{R_{zz}(\theta)} & \gate{U_b^\dagger} & \qw \\
        & \gate{U_a} & \qw & \gate{U_a^\dagger} & \qw
        \end{quantikz}
    }}
\end{equation*}
Here, $U_a$ and $U_b$ are single-qubit gates that map the $a$- and $b$-axes to the $z$-axis. Specifically, $U_x = H$ (Hadamard gate), $U_y = R_x\left(\frac{\pi}{2}\right)$ (rotation around the $x$-axis by $\frac{\pi}{2}$), and $U_z = I$ (identity gate).
\end{itemize}

In the case of RLVQC Global, the qubit indices range from $0$ to $n-1$, where $n$ is the total number of qubits. In contrast, RLVQC Block operates on a fixed 2-qubit system ($n = 2$).

All the gates within the action set $\mathcal{A}$ are chosen because they reduce to the identity operator when the corresponding parameter $\theta$ is set to zero. This property ensures that new gates can be appended to the ansatz without immediately altering the circuit's output state, provided they are initialized with $\theta = 0$. Therefore, the addition of gates followed by classical optimization throughout the training process is guaranteed to either maintain or improve the circuit's expectation value. This strategy is inspired by other methodologies established in the literature~\citep{rattew_2020, bilkis_2023}. Finally, for computing the circuit depth $d_t$ in the reward function (\ref{eq:reward}), we use the basis gate set $\{H, R_x, R_y, R_z, R_{zz}\}$.

\subsubsection{Episode Termination Condition}

As previously defined in Section \ref{sec:rlvqc:episodes}, the episode where the circuit is constructed terminates when either the reward fails to improve over successive steps or a fixed maximum number of steps is reached.

The first termination condition depends on the improvement of the reward, which is governed by the \textit{patience} hyperparameter, initially set to 3. If the step produces a circuit with a reward that is lower than the maximum obtained in the current episode, the patience is lowered by 1. On the other hand, if the reward improves the patience is increased by 1. The value of the patience is capped between 0 and its starting value, \idest 3. Once the patience reaches 0 the episode terminates. This mechanism prevents the agent from performing unnecessary actions that are unlikely to yield better outcomes.

The maximum number of gates which can be inserted into the circuit is governed by a hyperparameter and is set depending on the algorithm variant: for the Global one it corresponds to the number of gates contained in the circuit of a QAOA with $p=1$ layers, while for the Block variant it corresponds to a QAOA with $p=5$ layers. This difference is meant to compensate how the Block variant only has very few actions available within a block, despite the overall circuit being deeper.

\subsubsection{Parameter Optimization and Fine-tuning}
\label{sec:exp_protocol:effectiveness_block:param_opt}

To reduce the computational cost of gate parameter optimization in the inner stage and minimize the number of circuit executions required to estimate the cost function during training, the number of iterations performed by the classical optimizer during the episodes is limited to 50.
Numerical simulations are conducted using Qiskit QASM Simulator \revision{with 1000 circuit executions (shots)}, while circuit parameters are optimized via the COBYLA algorithm\footnote{We use the SciPy implementation of COBYLA, with all default hyperparameters. Documentation is available at: \url{https://docs.scipy.org/doc/scipy/reference/optimize.minimize-cobyla.html}.}.
COBYLA is selected for its recognized efficacy in noise-free environments and its high computational efficiency \citep{singh_2023, fernandez-pendas_2020}.
All COBYLA hyperparameters are set to their default values, with the exception of the number of iterations.

After the agent training has concluded and the best circuit or block has been identified (see Section ~\ref{sec:rlvqc:training_and_final}), the circuit parameters are further refined through a fine-tuning phase using COBYLA but with a larger iteration budget of 1000\footnote{This value ensures a limit in the number of iterations, but is typically not reached, as COBYLA always converges earlier.}. For the fine-tuning step, we initialize the parameters using the optimal parameters from the circuit that achieved the highest reward, providing a good starting point and facilitating convergence during fine-tuning. Note that this step follows the same parameter-sharing scheme adopted during training.
This two-phase approach limiting the number of optimization iterations during circuit construction and performing a final fine-tuning step reduces computational overhead during the exploratory phase while ensuring that the best identified circuit structure is optimized to its full potential.

\subsubsection{Hyperparameter Optimization}

The first stage of the experimental protocol for the first experiment involves setting the hyperparameters for both RLVQC and QAOA.
The choice of these hyperparameters is explained in the following sections. \revision{The full list of hyperparameters as well as additional details on the implementation are reported in Appendix \ref{app:implementation_hyperparameters}}.

Hyperparameter optimization is performed exclusively on the smallest problem instances, specifically the QUBO problems with $n = 8$ qubits and then the optimal hyperparameters are used for the problem instances of the same type and underlining graph topology but different size. 
For each problem instance, we perform five independent runs using the optimal hyperparameters.

\paragraph{RLVQC Global}

For optimizing the hyperparameters of RLVQC Global, we adopt a Bayesian optimization strategy, exploring a total of 50 hyperparameter configurations.\footnote{We rely on \texttt{gp\_minimize} provided as part of the \texttt{scikit-optimize} package, using the default acquisition strategy \texttt{gp\_hedge}} The configuration that produces the circuit with the best reward is selected.

The relevant hyperparameters in RLVQC Global correspond to the standard hyperparameters of the PPO algorithm. The penalty coefficient $\beta$ is a way to control the trade-off and is part of the experimental design to encourage low-depth circuits. Note that this coefficient is not optimized, and indeed it could not be optimized by maximising the reward, as its optimal value would likely be zero. \revision{To explore the trade-off between the Hamiltonian expectation value and circuit depth, one would need to train multiple instances of the agent with different values of $\beta$. Since our focus in this study is instead on modular ansatzes, where circuit depth can be more directly controlled by constraining the episode length, we fix a single value, $\beta = 0.1$, chosen so that its contribution to the reward remains comparable in magnitude to that of the expectation-value term.} 

We based our agent on OpenAI \texttt{Spinning Up} PPO implementation, and we adopt most of its default configurations. However, we optimize a selected subset of key hyperparameters to better tailor the algorithm to our setting, we summarize them in Table~\ref{tab:hyperparams_RLVQC-Global} along with their respective ranges and prior distributions.

\begin{table*}[ht]
\centering
\footnotesize
\caption{Ranges and prior distributions of hyperparameters for the RLVQC Global method.}
\renewcommand{\arraystretch}{1.3}
\begin{tabular}{>{\raggedright\arraybackslash}m{6.5cm} c c}
\toprule
\textbf{Hyperparameter} & \textbf{Range} & \textbf{Prior Distribution} \\ 
\midrule
Total number of steps (\texttt{total\_steps}) & 3000 & - \\
Number of steps per epoch (\texttt{steps\_per\_epoch}) & [100, 600] & Uniform \\ 
Maximum number of steps per episode (\texttt{max\_ep\_len}) & $2n$ & - \\ 
Policy Net learning rate (\texttt{pi\_lr}) & [$5 \cdot 10^{-6}$, $3 \cdot 10^{-3}$] & Log-uniform \\ 
Value Net learning rate (\texttt{vf\_lr}) & [$5 \cdot 10^{-6}$, $3 \cdot 10^{-3}$] & Log-uniform \\ 
Policy Net maximum number of gradient steps (\texttt{train\_pi\_iters}) & [4, 4096] & Uniform \\ 
Value Net maximum number of gradient steps (\texttt{train\_v\_iters}) & [4, 4096] & Uniform \\ 
\bottomrule
\end{tabular}
\label{tab:hyperparams_RLVQC-Global}
\end{table*}

The search ranges for \texttt{pi\_lr}, \texttt{vf\_lr}, \texttt{train\_pi\_iters}, and \texttt{train\_v\_iters} follow established best practices\footnote{See \url{https://medium.com/aureliantactics/ppo-hyperparameters-and-ranges-6fc2d29bccbe}}. The values of \texttt{total\_steps}\footnote{Unlike the original OpenAI \texttt{Spinning Up} PPO implementation, which specifies a fixed number of epochs, our setup defines a training budget in terms of total agent-environment interaction steps, from which the number of epochs is derived.} and the range for \texttt{steps\_per\_epoch} are chosen to ensure an adequate exploration of the circuit space and a sufficient number of training epochs.
The value of \texttt{max\_ep\_len}, which governs one termination condition of the episodes (see Section~\ref{sec:rlvqc}), is set to $2n$ because it is the number of steps required to obtain the QAOA circuit with depth $p=1$.
The prior distributions used during hyperparameter optimization are as follows: log-uniform priors are used for parameters that span several orders of magnitude, while uniform priors are used for the remaining hyperparameters.

\paragraph{RLVQC Block}

Optimizing the hyperparameters of RLVQC Global relative to the PPO algorithm had little effect on improving the algorithm's effectiveness. Therefore, for RLVQC Block we adopt the default PPO hyperparameters from the original PPO implementation with the exception of \texttt{total\_steps}, \texttt{steps\_per\_epoch} and \texttt{max\_ep\_len} which we choose to ensure a sufficient number of training epochs and the construction of an adequate number of circuits per epoch. The full set of PPO hyperparameters for RLVQC Block, is reported in Table~\ref{tab:hyperparams_RLVQC-Block}. 
The penalty coefficient $\beta$ is set to $0.1$ divided by the number of interacting qubits. This ensures that each action contributes a penalty to the reward comparable to that of RLVQC Global on account for how the block is repeated on every pair of interacting qubits.

\begin{table*}[ht]
\centering
\footnotesize
\caption{Values of hyperparameters for the RLVQC Block method in Experiment 1.}
\renewcommand{\arraystretch}{1.3}
\begin{tabular}{>{\raggedright\arraybackslash}m{10.5cm} c}
\toprule
\textbf{Hyperparameter} & \textbf{Value} \\ 
\midrule
Total number of steps (\texttt{total\_steps}) & 250 \\
Number of steps per epoch (\texttt{steps\_per\_epoch}) & 25 \\ 
Maximum number of steps per episode (\texttt{max\_ep\_len}) & 5  \\ 
Policy Net learning rate (\texttt{pi\_lr}) & $3\cdot 10^{-4}$ \\ 
Value Net learning rate (\texttt{vf\_lr}) & $10^{-3}$ \\ 
Policy Net maximum number of gradient steps (\texttt{train\_pi\_iters}) & 80 \\ 
Value Net maximum number of gradient steps (\texttt{train\_v\_iters}) & 80\\ 
\bottomrule
\end{tabular}
\label{tab:hyperparams_RLVQC-Block}
\end{table*}

\paragraph{QAOA}

For QAOA, the only hyperparameter that needs to be optimized is the circuit depth $p$. This hyperparameter is varied between 1 and 10, with the values to test selected from a uniform prior distribution.
Although the total number of possible configurations is 10 (corresponding to the possible values of $p$), a total of 50 configurations are tested. This is because each run involves inherent stochasticity, and we aim to better explore the impact of this variability across multiple trials. In QAOA, since the circuit structure is predefined only the gate parameters need to be optimized, therefore we optimize them with COBYLA directly with 1000 iterations.

\subsection{Experiment 2: Extending the Modular Ansatz to Larger Instances}
\label{sec:exp_protocol:block_transfer}
\revision{
This section describes the second experiment, which evaluates whether \textit{RLVQC Block} can discover circuit blocks that remain effective when used to compose an ansatz for larger problem instances. The primary objective is to test the core premise of our methodology: blocks identified during a small-scale \emph{structure discovery} phase can be used to instantiate, via a prescribed composition rule, an ansatz suitable for a VQE-style optimization procedure to tackle more complex instances. Demonstrating that such a modular ansatz can be extended would support the idea that classical learning can be leveraged where simulation is feasible, while the resulting circuit structure can be deployed in regimes where classical simulation is challenging or impractical.

The experiment is structured in two parts. First, we train the \textit{RLVQC Block} parameter-sharing variants introduced in Section~\ref{sec:rlvqc:param_sharing} to identify optimized modular gate sequences on small instances. Second, we compose the final ansatzes as described in Section~\ref{sec:rlvqc:training_and_final}, and evaluate their solution quality on larger problem instances. In order to assess the variance we repeat each experiment 5 times.
}

\subsubsection{Search for Modular Blocks}
\revision{
When employing RLVQC Block to identify modular structures that would enable the extension of the ansatzes to larger problems, we introduce specific modifications to the experimental protocol established in the previous experiment. These adjustments reflect the shifted focus toward discovering a block that can be deployed to larger problem instances.

We evaluate all three RLVQC Block parameter-sharing variants defined in Section~\ref{sec:rlvqc:param_sharing} to analyze the trade-off they offer. For this search phase, the models are trained on all the problem instances with $n=8$ qubits described in Section~\ref{sec:exp_protocol:instances}.

To ensure the discovered blocks maintain low circuit depth when the final ansatz is created, the action space is restricted to fundamental rotation gates and the $CX$ gate:
\begin{equation}
\label{eq:gate_set_exp2}
    \mathcal{A} \coloneq \{R_X(\theta), R_Y(\theta), R_Z(\theta), CX\}.
\end{equation}
Within this framework, the penalty coefficient for circuit depth in the reward function, as defined in (\ref{eq:reward}), is set to zero to prioritize the optimization of the solution quality, note however that the depth of the ansatz can be easily controlled by constraining the episode length.

During training, each episode utilizes the COBYLA optimizer with a limit of 50 iterations for parameter tuning.
An episode terminates once exactly three gates have been inserted into the block, a constraint chosen to maintain consistency with the foundational QAOA block, which comprises two $CX$ gates and one $R_Z$ gate.

RLVQC uses the hyperparameters detailed in Table~\ref{tab:hyperparams_RLVQC-Block-exp2}. The full list of hyperparameters as well as additional details on the implementation are reported in Appendix \ref{app:implementation_hyperparameters}.
Note that, one of the benefits of splitting QAS in two phases, is that during the discovery phase we can perform a more thorough optimization of the ansatz, because that is not tied to the specific problem instance we aim to solve, therefore we give the agent a budget of 3000 interaction steps for completing the training process but constrain episodes to a more limited 3 steps.

After training, the circuit achieving the highest approximation ratio throughout the entire search is selected, and its underlying block structure is used to extend the ansatz to larger problems.
Notably, the gate parameters are not subjected to a final fine-tuning phase. This approach ensures that the discovered blocks are easily optimizable, even when computational resources for classical optimization are limited.
}

\begin{table*}[ht]
\centering
\footnotesize
\caption{Values of hyperparameters for the RLVQC Block method in Experiment 2.}
\renewcommand{\arraystretch}{1.3}
\begin{tabular}{>{\raggedright\arraybackslash}m{10.5cm} c}
\toprule
\textbf{Hyperparameter} & \textbf{Value} \\ 
\midrule
Total number of steps (\texttt{total\_steps}) & 3000 \\
Number of steps per epoch (\texttt{steps\_per\_epoch}) & 30 \\ 
Maximum number of steps per episode (\texttt{max\_ep\_len}) & 3  \\ 
Policy Net learning rate (\texttt{pi\_lr}) & $3\cdot 10^{-4}$ \\ 
Value Net learning rate (\texttt{vf\_lr}) & $10^{-3}$ \\ 
Policy Net maximum number of gradient steps (\texttt{train\_pi\_iters}) & 80 \\ 
Value Net maximum number of gradient steps (\texttt{train\_v\_iters}) & 80\\ 
\bottomrule
\end{tabular}
\label{tab:hyperparams_RLVQC-Block-exp2}
\end{table*}

\subsubsection{Larger Scale Analysis}
\revision{
Following the identification of the optimal blocks for $n=8$, we evaluate their effectiveness by testing them on all problem instances defined in Section~\ref{sec:exp_protocol:instances}, in particular on the larger instances with $n=12$ and $n=16$ qubits. The procedure used to construct the final ansatz from the learned block is the same described in Section~\ref{sec:rlvqc:training_and_final}. We also include standard QAOA and ma-QAOA as comparative baselines.

A central goal of this analysis is to assess whether the effectiveness of the resulting modular ansatzes remains stable as the problem size increases. Since the blocks are discovered on smaller instances, which are typically simpler, we evaluate whether composing the same learned block on larger instances preserves solution quality, or whether its effectiveness degrades as the dimensionality of the problem grows. This experiment therefore quantifies the stability of the learned modular structure under this size shift. For this reason, we learn the blocks on the smaller $n=8$ problems and we then focus our analysis on their effectiveness for the largest problem instances $n=16$.

For both the proposed RLVQC methods and the baselines, the overall circuit depth is controlled by the hyperparameter $p$, representing the number of repeated layers. We evaluate the effectiveness of algorithms and baselines across a range of $p \in \{1, 2, 3, 4\}$.
All numerical simulations are performed using the Qiskit QASM Simulator with $n_\text{runs}=1000$ shots.
The classical optimization of gate parameters is conducted using COBYLA, with a maximum budget of 1000 iterations.
}

\section{Results}
\label{sec:results}

This section reports the empirical results of our study on modular ansatz discovery and extension to larger instances. Section~\ref{sec:results:effectiveness_block} evaluates the \emph{structure discovery} phase by comparing \textit{RLVQC Block} with \textit{RLVQC Global} and the QAOA baseline, assessing whether imposing a block structure affects solution quality and circuit resources. Section~\ref{sec:results:block_transfer} then evaluates the \emph{deployment} phase, analyzing whether blocks discovered on small-scale instances remain effective when instantiated to solve larger problem instances. \revision{Note that we report the results of a significant number of experiments on a total of 24 problem instances. Experiment 1, on the effectiveness of the block structure, required 3'400 experiments, due to the hyperparameter optimization and the 5 independent executions, while Experiment 2, on extendability to larger problems, which includes the parameter-sharing schemes and different depths $p$, required 7'200 experiments.}

\subsection{Experiment 1: Effectiveness of the Block Structure}
\label{sec:results:effectiveness_block}
\revision{
In this section, we report the results of the first experiment which assesses the impact of imposing a modular block structure during the \emph{structure discovery} phase. Following the protocol in Section~\ref{sec:exp_protocol}, we compare \textit{RLVQC Block} with the unconstrained \textit{RLVQC Global} variant and the standard QAOA baseline, with the specific goal of evaluating whether restricting the action space through a block-based construction hinders learning. The results indicate that this structural restriction is not detrimental in our setting, and that the block-based approach yields competitive, and in several cases improved, solution quality and circuit characteristics.

We first discuss solution quality in terms of approximation ratio, and then analyze the discovered circuits in terms of total gate count, circuit depth, and gate composition. For the sake of brevity, we focus on results for $n=16$ qubits, as they are the most significant. Results on smaller instances ($n=8$ and $n=12$) show similar trends and are available in Appendix~\ref{app:effectiveness_block}.
}

\subsubsection{Approximation Ratio}
We begin by analyzing solution quality, and report the mean and standard deviation of the approximation ratios achieved by the three algorithms over five independent executions across the $n=16$ problem instances in Table~\ref{tab:ar_n=16}.

\begin{table*}[ht]
    \centering
    \footnotesize
    \caption{
    Approximation ratios achieved by RLVQC Global and RLVQC Block compared to the QAOA baseline on QUBO problem instances with $n = 16$ qubits, evaluated across various graph topologies. Each value reports the mean and standard deviation over five independent runs. For each instance, RLVQC results are typeset in \textbf{bold} when they outperform QAOA, and the highest average approximation ratio is \underline{underlined}.
    \label{tab:ar_n=16}
    }
    %\resizebox{\textwidth}{!}{
    \begin{tabular}{rr|ccccc}
    \toprule
        \textbf{Problem} & \textbf{Topology} & \textbf{QAOA} & \textbf{RLVQC Global} & \textbf{RLVQC Block} \\ 
        \midrule
        \multirow{8}{*}{\begin{tabular}{@{}c@{}}Maximum \\ Cut\end{tabular}} 
        & 2d-grid - 4 & 0.693 ± 0.051 & 0.576 ± 0.042 & \underline{\textbf{0.994 ± 0.004}} \\ 
        & 3-reg & 0.813 ± 0.027 & 0.640 ± 0.021 & \underline{\textbf{0.942 ± 0.030}} \\ 
        & barabási-albert - 4 & 0.752 ± 0.024 & \textbf{0.767 ± 0.023} & \underline{\textbf{0.997 ± 0.002}} \\ 
        & barabási-albert - 8 & 0.754 ± 0.011 & \textbf{0.788 ± 0.010} & \underline{\textbf{0.997 ± 0.002}} \\ 
        & cycle & 0.838 ± 0.016 & 0.585 ± 0.036 & \underline{\textbf{0.998 ± 0.001}} \\ 
        & erdős-rényi - 0.2 & 0.845 ± 0.070 & 0.727 ± 0.039 & \underline{\textbf{0.915 ± 0.007}} \\ 
        & erdős-rényi - 0.7 & 0.772 ± 0.002 & \textbf{0.820 ± 0.017} & \underline{\textbf{0.945 ± 0.032}} \\ 
        & star & 0.992 ± 0.002 & 0.609 ± 0.025 & \underline{\textbf{0.996 ± 0.007}} \\ 
        \midrule
        \multirow{8}{*}{\begin{tabular}{@{}c@{}}Maximum \\ Clique\end{tabular}} 
        & 2d-grid - 4 & 0.764 ± 0.001 & \underline{\textbf{0.910 ± 0.015}} & \textbf{0.832 ± 0.060} \\ 
        & 3-reg & 0.764 ± 0.001 & \underline{\textbf{0.912 ± 0.009}} & \textbf{0.887 ± 0.056} \\ 
        & barabási-albert - 4 & 0.765 ± 0.002 & \textbf{0.930 ± 0.020} & \underline{\textbf{0.936 ± 0.027}} \\ 
        & barabási-albert - 8 & 0.785 ± 0.022 & \underline{\textbf{0.928 ± 0.017}} & \textbf{0.888 ± 0.056} \\ 
        & cycle & 0.765 ± e-04 & \underline{\textbf{0.910 ± 0.011}} & \textbf{0.868 ± 0.061} \\ 
        & erdős-rényi - 0.2 & 0.764 ± 0.001 & \underline{\textbf{0.914 ± 0.022}} & \textbf{0.858 ± 0.058} \\ 
        & erdős-rényi - 0.7 & 0.779 ± 0.007 & \textbf{0.919 ± 0.013} & \underline{\textbf{0.974 ± 0.013}} \\ 
        & star & 0.765 ± 0.001 & \underline{\textbf{0.928 ± 0.019}} & \textbf{0.823 ± 0.081} \\ 
        \midrule        
        \multirow{8}{*}{\begin{tabular}{@{}c@{}}Minimum \\ Vertex \\ Cover\end{tabular}}
        & 2d-grid - 4 & 0.938 ± 0.038 & \textbf{0.941 ± 0.011} & \underline{\textbf{0.994 ± 0.005}} \\ 
        & 3-reg & 0.975 ± 0.004 & 0.935 ± 0.015 & \underline{\textbf{0.993 ± 0.001}} \\ 
        & barabási-albert - 4 & 0.785 ± 0.017 & \textbf{0.948 ± 0.002} & \underline{\textbf{0.992 ± 0.001}} \\ 
        & barabási-albert - 8 & 0.769 ± 0.006 & \textbf{0.934 ± 0.007} & \underline{\textbf{0.992 ± 0.001}} \\ 
        & cycle & 0.967 ± 0.006 & 0.943 ± 0.011 & \underline{\textbf{0.991 ± 0.007}} \\ 
        & erdős-rényi - 0.2 & 0.959 ± 0.027 & 0.952 ± 0.020 & \underline{\textbf{0.992 ± 0.002}} \\ 
        & erdős-rényi - 0.7 & 0.764 ± 0.001 & \textbf{0.924 ± 0.006} & \underline{\textbf{0.997 ± 0.001}} \\ 
        & star & 0.937 ± 0.016 & \textbf{0.975 ± 0.001} & \underline{\textbf{0.997 ± 0.002}} \\ 
        \bottomrule
    \end{tabular}
    %}
\end{table*}

\revision{
Overall, the results indicate that introducing a modular block structure does not make learning harder. In most instances, \textit{RLVQC Block} achieves higher approximation ratios than \textit{RLVQC Global}, showing that the restriction to blocks is not detrimental and can in fact be beneficial. Specifically, \textit{Block} is outperformed by \textit{Global} only on six instances of Maximum Clique. Both RLVQC variants are also competitive with the standard QAOA baseline on these problems, even though this comparison is not the focus of our study. The \textit{Global} variant is sometimes inferior to QAOA on Maximum Cut, and only very rarely on Minimum Vertex Cover. In contrast, \textit{Block} is the best performing method overall, consistently achieving the highest approximation ratios, always above 0.8, mostly above 0.9, and occasionally approaching 1.
We stress that our goal is not to claim superiority over other QAOA variants or over classical solvers, but to assess the effect of modularity during structure discovery and to support its extendability across problem sizes.
}

A notable aspect of these results is that the \textit{Block} variant achieves the best solution quality despite operating with the most constrained action space. Even with a simple agent and default hyperparameters, the modular construction consistently yields strong results, and substantially improves over ad hoc heuristic circuit choices such as standard QAOA. This suggests that, even in a relatively simple form, RL provides a powerful tool for the automated discovery of modular ansatzes.

We can present two possible explanations of why RLVQC Global does not achieve the same result quality as RLVQC Block. First, despite having a larger number of available actions, the Global variant is designed to construct circuits with a more limited maximum depth compared to the Block variant. It is possible that reducing the number of available actions but allowing deeper circuits is a more effective trade-off. Allowing deeper circuits could potentially improve the quality of the results, however this would significantly increase the computational cost, and it is uncertain whether this would lead to better results. In fact, limiting circuit depth can sometimes be advantageous by simplifying the training process. The second possible reason for the limited effectiveness of the Global variant is that removing the structural requirement that actions should only affect the interacting qubit pairs removes some implicit problem-related information and creates a much larger action space which is more challenging for the agent to explore effectively. This again could likely be addressed with a more careful training and hyperparameter optimization process, or with a more expressive agent architecture. However, this would come again at the cost of an increased computational load. Among the future research directions could be to explore how to provide the agent with more information on the problem it is tasked to solve while maintaining high flexibility in designing the circuit.

\revision{
In summary, this experiment supports the main premise of our approach: restricting circuit discovery to a modular, compositional form learned with RL does not reduce effectiveness and can instead improve it. In particular, the \textit{Block} variant learns a reusable two-qubit building block, which constrains the search to circuits that reflect the local interaction structure of the problem. This motivates the second part of our study, where we test whether the same learned modular structure discovered on small instances can be extended to construct effective ansatzes for larger problem sizes.
}

\subsubsection{Circuit Composition}

\begin{figure}[ht]
    \centering    
    \includegraphics[width=.49\textwidth]{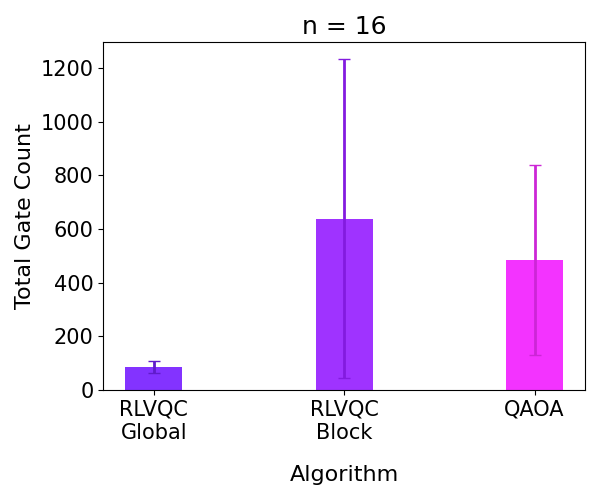}\hfill
    \includegraphics[width=.49\textwidth]{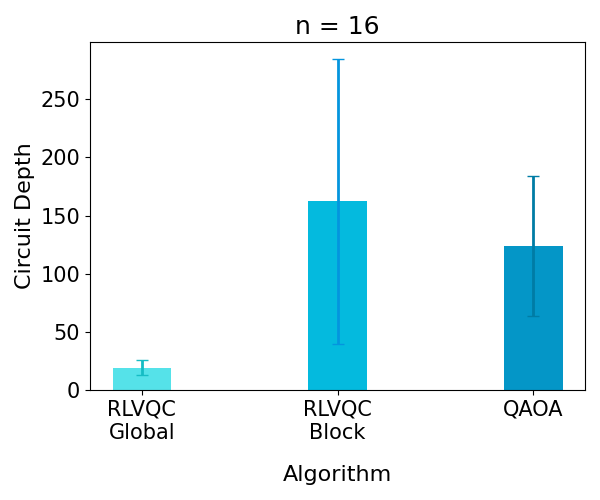}
    \caption{Comparison of the average total gate count (left) and circuit depth (right) of the optimal circuits obtained using RLVQC Global, RLVQC Block, and the baseline QAOA, across instances with $n=16$. Each histogram corresponds to a different algorithm. The bar above each histogram represents the standard deviation. \label{fig:gate_count_and_depth_qubo_n=16}} 
\end{figure}

\begin{figure}[ht]
    \centering
    \includegraphics[width=0.88\textwidth]{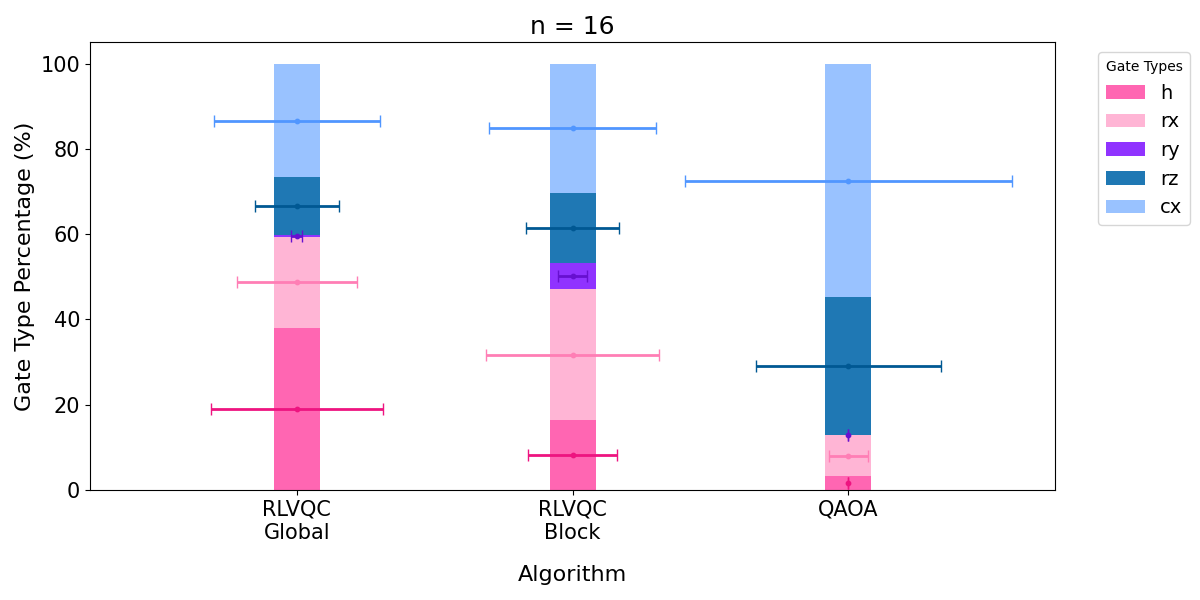}
    \caption{
    Average gate usage by algorithm, expressed as a percentage of the total gate count in the optimal circuits produced by RLVQC Global, RLVQC Block, and the baseline QAOA, for , across instances with $n=16$. Each histogram corresponds to a different algorithm. Standard deviation bars are proportional to the standard deviation within each algorithm.
    \label{fig:gate_usage_qubo_n=16}
    }
\end{figure}

We now analyze the structural properties of the circuits produced by each algorithm, focusing on two key aspects: the total gate count and circuit depth, presented in Fig.~\ref{fig:gate_count_and_depth_qubo_n=16}, and the distribution of specific gate types, shown in Fig.~\ref{fig:gate_usage_qubo_n=16}.

Both analyses are based on the circuits without any simplification, expressed in terms of the gates in the basis \{$H$, $R_x$, $R_y$, $R_z$, $CX$\}.

\paragraph{Gate Count and Circuit Depth}

Fig.~\ref{fig:gate_count_and_depth_qubo_n=16} presents a comparison of the average number of gates and circuit depth across the three algorithms for the instances with $n=16$. Error bars indicate the standard deviation, calculated over five independent runs.

The circuits generated by RLVQC Global are the shortest among the algorithms tested, which is a particularly positive outcome. This indicates that RLVQC Global is able to construct more efficient circuits with fewer gates, which is advantageous for both computational cost and reducing the likelihood of errors, especially in noisy hardware environments.

In the case of the Block variant, the gate count and circuit depth are comparable to those of QAOA, considering the variance.
RLVQC Block constructs deeper circuits than RLVQC Global, likely because it is designed to generate circuits with a maximal depth corresponding to a 5-layer QAOA, whereas RLVQC Global constructs circuits with a depth corresponding to a 1-layer QAOA (see Section~\ref{sec:exp_protocol:effectiveness_block:param_opt}).

Moreover, we highlight that the penalty coefficient in the reward function can be adjusted to balance the trade-off between circuit complexity and solution quality. By modifying this penalty coefficient, the circuit depth can be controlled, allowing for better alignment with the specific requirements. 
Therefore, further optimizations are possible by either reducing the step budget allocation or increasing the penalty coefficient. The current setup offers a good balance, enabling the construction of circuits with depths and gate counts comparable to QAOA, while achieving superior results. However, with adjustments to the reward function or by tightening the gate count budgets, even more efficient circuits could be achieved.

Finally, the results for $n=8$ and $n=12$ presented in Appendix~\ref{app:effectiveness_block} (see Fig.~\ref{fig:gate_count_and_depth_n=8_12}) indicate that gate count and circuit depth follow similar trends across all evaluated algorithms. This highlights a consistent relationship between the total number of gates and the overall complexity of the synthesized circuits.

\paragraph{Gate Usage}

Fig.~\ref{fig:gate_usage_qubo_n=16} presents the percentage distribution of gate types used in the circuits generated by each algorithm.
The figure also provides an indication of the standard deviations. Although the exact numerical values are not explicitly reported, the bars are proportional to the standard deviation within each algorithm. Specifically, the total length of the bars is fixed across all algorithms, and the relative lengths reflect the variability of the standard deviations across different gate types within each algorithm.

The use of $R_x$ and $R_z$ gates is modest across all circuits, with $R_y$ gates being notably absent in QAOA by definition. In RLVQC Global, $R_y$ gates are almost negligible, while in RLVQC Block, they are more prevalent, though still relatively limited.

One of the most significant observations is that the number of $CX$ gates in both RLVQC Block and Global is much lower than in QAOA. This is advantageous, as 2-qubit gates, such as $CX$, are generally more difficult to implement, especially on real hardware, and are more prone to introducing noise. Consequently, reducing the use of $CX$ gates helps in minimizing potential hardware issues, maintaining higher fidelity in computations, particularly in noisy environments. Naturally, a sufficient number of $CX$ is important to ensure the circuit generates a sufficient level of entanglement. These results also highlight how learning-based methods such as the one proposed here can be effective in identifying alternative ansatz compositions that achieve comparable, or even improved, solution quality while exhibiting different gate compositions, which may be better suited to specific hardware constraints or design requirements.

Furthermore, as illustrated by the circuit compositions for $n=8$ and $n=12$ in Appendix~\ref{app:effectiveness_block} (see Fig.~\ref{fig:gate_usage_qubo_n=8_12}), the gate distribution remains consistent within each algorithm across varying problem sizes. This observation suggests that the structural composition of the synthesized circuits is invariant to the system size, regardless of the number of qubits $n$.

\subsection{Experiment 2: Extending the Modular Ansatz to Larger Instances}
\label{sec:results:block_transfer}
\revision{
This section reports the results of the second experiment, which evaluates the effectiveness of the modular blocks identified during the structure discovery phase, when used to build ansatzes for larger problems. Building on the previous experiment, where we showed that the block-based construction is effective and can be beneficial, we now test the central goal of our methodology: separating discovery from deployment by learning blocks on small instances and extending the modular ansatz to larger instances using the discovered block and the prescribed composition rule. We analyze how solution quality varies with problem size and with the number of circuit parameters for the different \textit{Block} parameter-sharing variants and baselines. We also compare the block-based constructions against the baselines to assess whether observed differences are statistically significant.
}

\subsubsection{Quality Assessment Results}

\begin{figure}[ht]
    \centering
    \includegraphics[width=\textwidth]{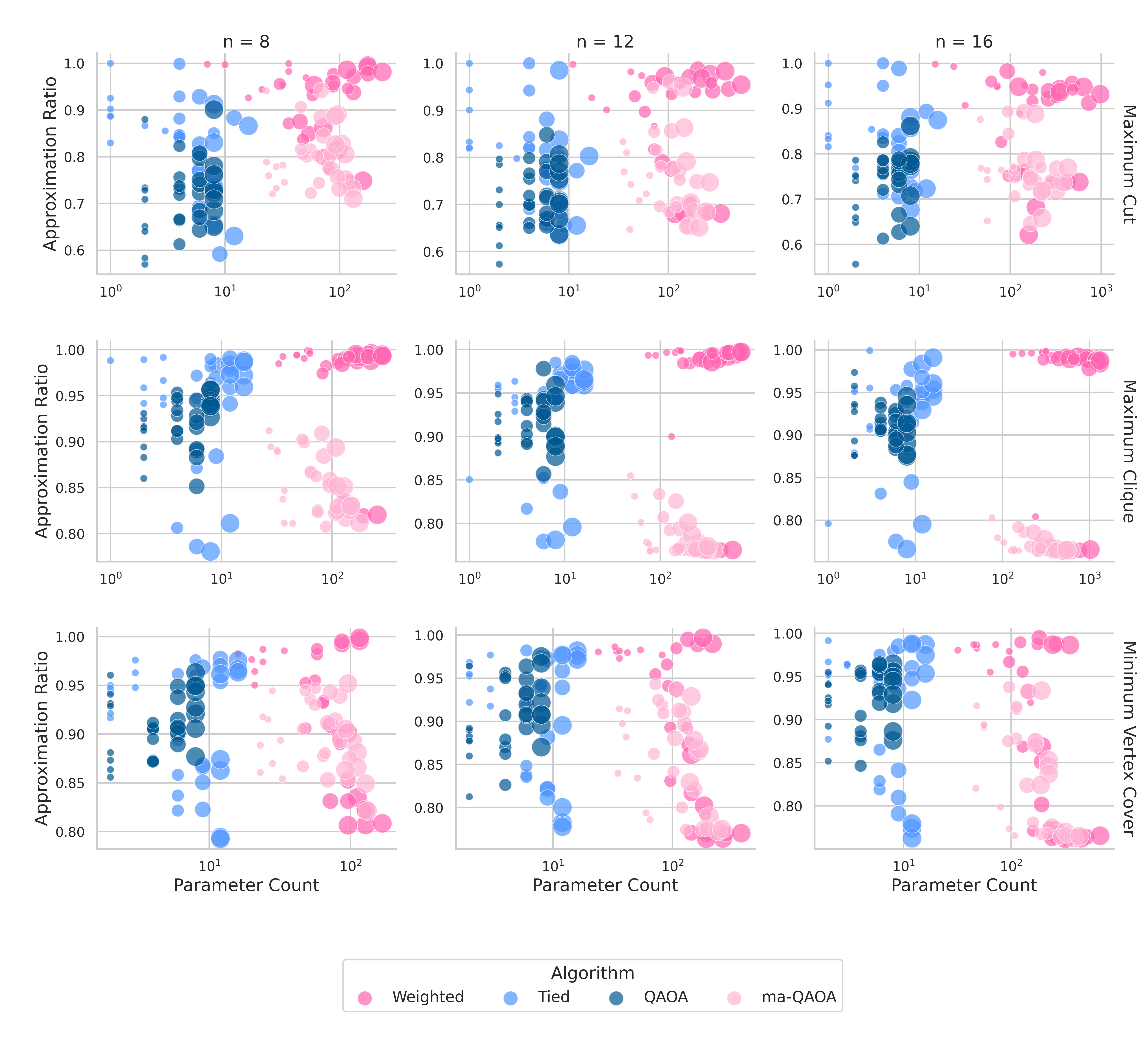}
    \caption{\revision{Average approximation ratio as a function of the total circuit parameter count across all benchmarked instances. Each marker represents a specific combination of problem instance, algorithm, and circuit depth $p$. The grid arrangement categorizes problem instances by size $n$ (horizontal axis) and problem type (vertical axis). Marker colours identify the specific algorithm employed, while the marker size is proportional to the number of layers $p$.}}
    \label{fig:approx_ratio_params_scatterplot}
\end{figure}

\revision{
Here, we provide a summary analysis of the average solution quality achieved by each algorithm and baseline in relation to the number of independent parameters in the circuit. This analysis is critical for evaluating the practical viability of the proposed methods, as achieving high-quality solutions must be balanced with resource efficiency to establish an optimal effectiveness-cost trade-off. Indeed, if ansatz allows to achieve very good approximation rations but only under extensive parameter optimization, a comparison that does not discuss this fact could be rather unfair. Note that this section provides a summary view of the trade-offs, while we will compare the results per each instance against the baselines, and assess the statistical significance of the results, in the following Section \ref{sec:results:block_transfer:statistical}.

A comprehensive overview of this trade-off is provided in Fig.~\ref{fig:approx_ratio_params_scatterplot}. The figure reports, for each problem instance and method, the achieved approximation ratio as a function of the number of circuit parameters, allowing a direct comparison between solution quality and parameter budget. Each point corresponds to the mean of the 5 executions on a specific problem instance, and points higher on the vertical axis indicate better solutions. By inspecting how the different methods populate the plot, we can assess whether some approaches achieve comparable solution quality with fewer parameters, and whether a consistent trend emerges as the problem size increases. Based on the underlying circuit definitions, the most pertinent comparisons between baselines and our parameter-sharing variants are between \texttt{Weighted} and ma-QAOA, and between \texttt{Tied} and QAOA, as these pairs utilize a comparable number of parameters.
For visual clarity, we exclude the results for the \texttt{Agnostic} variant, as its behaviour closely mirrors that of \texttt{Weighted}. 

Our first observation is that the distribution of data points, which represent solution quality relative to parameter count, remains consistent across all evaluated problem sizes $n$. In particular, the relative positioning of the methods is largely preserved as $n$ increases, and we do not observe a systematic drop in approximation ratios for the block-based ansatzes when moving from $n=8$ to $n=12$ and $n=16$, with only few exceptions. This stability suggests that the modular blocks discovered at $n=8$ retain their usefulness when reused at larger sizes, supporting the central idea of separating structure discovery from deployment.

Second, the methods naturally cluster into two groups based on parameter budget. The first group, which includes \texttt{Tied} and standard QAOA, operates with fewer independent parameters. The second group, which includes \texttt{Weighted} and ma-QAOA, uses a substantially larger number of parameters. This separation is useful because it shows that parameter sharing acts as a direct control knob for navigating the trade-off between solution quality and resource requirements.

A further observation is that the plot reveals a trade-off frontier: higher approximation ratios are frequently achieved at higher parameter counts, but with diminishing returns in many instances. In this context, QAOA and \texttt{Tied} are the most resource-efficient options, often reaching approximation ratios that are reasonably close to the higher-parameter methods while requiring far fewer parameters. Comparing these two, \texttt{Tied} generally achieves higher approximation ratios than standard QAOA on Maximum Cut, while exhibiting a broader spread on Maximum Clique and Minimum Vertex Cover, indicating that reuse of a learned block can help but may be more instance-dependent for some tasks. 

Finally, while \texttt{Weighted} and ma-QAOA can reach higher approximation ratios than their low-parameter counterparts, the marginal gains in solution quality may not always justify the substantial increase in computational resources. Notably, \texttt{Weighted} exhibits good results more consistently than ma-QAOA.

Table~\ref{tab:cobyla_iterations} provides a complementary view by reporting the average number of COBYLA iterations required for convergence during parameter tuning, with a hard cap of 1000 iterations. A clear pattern is that \texttt{Tied} and QAOA converge in very few iterations (typically a few tens to low hundreds). This supports the interpretation that the strong solution quality of \texttt{Tied} is not driven by an extensive parameter optimization for each instance, but rather by the effectiveness of the learned modular structure and its parameter-sharing schemes. Remember that the discovery phase, as described in Section \ref{sec:exp_protocol:block_transfer}, used only 50 COBYLA iterations, which may have created a bias towards ansatzes that reach reasonably good results quickly but that may be more difficult to fine-tune. Performing the discovery phase with a larger number of COBYLA iterations may allow to control this behaviour, possibly by allowing to discover ansatzes able to reach better results when extensive parameter optimization is available. By contrast, ma-QAOA, \texttt{Agnostic}, and \texttt{Weighted} often saturate the iteration budget, which is consistent with a more demanding fine-tuning step due to a significantly larger number of independent parameters. From a trade-off perspective, while highly parameter intensive ansatzes may allow to reach better approximation ratios, if this is achieved only at the cost of extensive parameter optimization the comparison between approximation ratios alone may be misleading.
}

\begin{table}[ht]
\centering
\caption{\revision{Average number of COBYLA iterations for the parameter optimization of the ansatzes on instances with $n=16$, across problem types and circuit depths $p$.}}
\label{tab:cobyla_iterations}
\begin{tabular}{cc|cc|ccc}
\toprule
\textbf{Problem} & \textbf{p} & \textbf{Tied} & \textbf{QAOA} & \textbf{Agnostic} & \textbf{Weighted} & \textbf{ma-QAOA} \\
\midrule
\multirow{4}{*}{\makecell[c]{Maximum \\ Cut}} &  1 & 24 ± 7 & 31 ± 1 & 621 ± 333 & 691 ± 340 & 672 ± 105 \\ 
         & 2 & 59 ± 13 & 50 ± 3 & 955 ± 85 & 973 ± 36 & 994 ± 14 \\ 
         & 3 & 79 ± 18 & 66 ± 5 & 997 ± 7 & 1000 ± 0 & 1000 ± 0 \\ 
         & 4 & 100 ± 27 & 88 ± 5 & 1000 ± 0 & 1000 ± 0 & 1000 ± 0 \\ 
\midrule
\multirow{4}{*}{\makecell[c]{Maximum \\ Clique}} & 1 & 35 ± 8 & 29 ± 1 & 939 ± 133 & 1000 ± 0 & 891 ± 138 \\ 
         & 2 & 75 ± 14 & 48 ± 1 & 1000 ± 0 & 1000 ± 0 & 1000 ± 0 \\ 
         & 3 & 109 ± 24 & 66 ± 2 & 1000 ± 0 & 1000 ± 0 & 1000 ± 0 \\ 
         & 4 & 140 ± 33 & 84 ± 3 & 1000 ± 0 & 1000 ± 0 & 1000 ± 0 \\ 
\midrule
\multirow{4}{*}{\makecell[c]{Minimum \\ Vertex \\ Cover}} & 1 & 35 ± 5 & 31 ± 1 & 610 ± 345 & 687 ± 233 & 638 ± 138 \\ 
         & 2 & 75 ± 11 & 49 ± 2 & 898 ± 142 & 1000 ± 0 & 978 ± 35 \\ 
         & 3 & 107 ± 16 & 68 ± 2 & 972 ± 69 & 1000 ± 0 & 1000 ± 0 \\ 
         & 4 & 137 ± 28 & 88 ± 2 & 1000 ± 0 & 1000 ± 0 & 1000 ± 0 \\ 
\bottomrule
\end{tabular}
\end{table}

\subsubsection{Statistical Significance of Solution Quality Improvements}
\label{sec:results:block_transfer:statistical}
\revision{
To evaluate whether the modular blocks discovered through our \textit{RLVQC Block} strategies remain effective as the problem size increases, we analyse the solution quality across $n \in \{8,12,16\}$ using approximation ratios. Since both the block discovery procedure and the subsequent parameter optimization are stochastic, observed differences across methods can arise from variability across runs. We therefore complement the analysis on larger problem sizes with a statistical significance study, testing whether the improvements observed relative to the QAOA and ma-QAOA baselines are robust and not explained by random fluctuations.
}

\begin{table}[ht]
\centering
\caption{\revision{Right-sided Wilcoxon signed-rank test results evaluating whether the proposed algorithms exhibit superior effectiveness compared to the baselines across diverse problem types and sizes $n$, with both configured at their optimal circuit depths $p$. Statistical outcomes are categorized as follows: \win denotes that the algorithm is significantly superior to the baseline; \winnosig indicates superior effectiveness that does not reach statistical significance; and \loss indicates that the algorithm is outperformed by the baseline.}}
\label{tab:wilcoxon_mean_test_best_p}
\begin{tabular}{ll|ccc|ccc}
\toprule
& & \multicolumn{3}{c|}{\textbf{QAOA}} & \multicolumn{3}{c}{\textbf{ma-QAOA}} \\
\textbf{Problem} & \textbf{Variant} & \textbf{n=8} & \textbf{n=12} & \textbf{n=16} & \textbf{n=8} & \textbf{n=12} & \textbf{n=16} \\
\midrule
\multirow{3}{*}{\makecell[c]{Maximum \\ Cut}} 
& \texttt{Agnostic} & \win & \win & \win &\win &\win &\win \\
& \texttt{Weighted} & \win & \win & \win &\win &\win &\win \\
& \texttt{Tied} & \win & \win & \win &\win &\win &\win \\
\midrule
\multirow{3}{*}{\makecell[c]{Maximum \\ Clique}} 
& \texttt{Agnostic} & \win & \win & \win &\win &\win &\win \\
& \texttt{Weighted} & \win & \win &\winnosig &\win &\win &\win \\
& \texttt{Tied} & \win &\winnosig &\loss &\win &\win &\win \\
\midrule
\multirow{3}{*}{\makecell[c]{Minimum \\ Vertex \\ Cover}} 
& \texttt{Agnostic} & \win & \win &\winnosig &\win &\win &\win \\
& \texttt{Weighted} & \win & \win & \win &\win &\win &\win \\
& \texttt{Tied} &\winnosig & \win &\winnosig &\win &\win &\win \\
\bottomrule
\end{tabular}
\end{table}

\revision{
For each algorithm, problem type, and $n$, we first identify the circuit depth $p \in \{1, \dots, 4\}$ that yields the highest mean approximation ratio, so that each method is evaluated under its best-performing configuration within the considered depth range. Using the results at these selected depths, we then perform pairwise comparisons between the \textit{RLVQC Block} variants and the baselines for each problem size $n$. We emphasize that this analysis is intended to assess whether the effectiveness of the learned modular structure remains stable and consistent as $n$ increases, rather than to claim overall superiority over alternative variational approaches or classical solvers.

We employ the Wilcoxon signed-rank test~\citep{wilcoxon_1992} to assess whether the \textit{RLVQC Block} variants achieve a significantly higher mean approximation ratio than the baselines. This non-parametric test was selected for two primary reasons: first, the distribution of approximation ratios does not necessarily satisfy the normality assumption required for a $t$-test; and second, the samples are naturally paired, as each algorithm is evaluated on the exact same set of problem instances.

The results are summarized in Table~\ref{tab:wilcoxon_mean_test_best_p}. Across the tested configurations, the \textit{RLVQC Block} variants are consistently superior to ma-QAOA according to the test. Relative to standard QAOA, they are usually statistically superior as well: only in 5 out of 27 cases the improvement does not reach statistical significance, and in 1 case the average approximation ratio is lower.

Finally, our analysis reveals a significant advantage in resource efficiency: the optimal approximation ratio for the RLVQC Block variants is typically achieved at lower circuit depths ($p\approx 1$) compared to QAOA and ma-QAOA. This indicates that the reinforcement learning agent discovers more efficient gate sequences.
}

\FloatBarrier

\section{Conclusions}
\label{sec:conclusions}
\revision{
We study the use of classical machine learning methods as support tools for quantum computing, focusing on the automated design of variational ansatzes. While data-driven approaches can be effective for discovering circuit structures, a key practical limitation is that learning can become impractical as the number of qubits increases, both due to the cost of simulation and due to the scaling of observation and optimization procedures. Motivated by this, we investigate a framework that separates \emph{structure discovery} from \emph{deployment}: an ansatz is learned in a small-scale regime where classical learning is feasible, and the resulting modular structure is then extended to construct circuits for larger problem instances.

To this end, we introduced two reinforcement learning based methods for variational circuit design, aimed at QUBO optimization problems. \textit{RLVQC Global} represents the fully unconstrained variant, where the agent sequentially places gates without a predefined architectural template. \textit{RLVQC Block} instead learns a regular modular structure by discovering a two-qubit block that is composed according to the interaction structure of the problem. This design is intentionally restrictive, as it enables a clear composition rule for deployment on larger instances.

Our experiments address two questions. First, we assess whether imposing this modular structure, and the corresponding reduction in action space, negatively affects learning. The results show the opposite: \textit{RLVQC Block} consistently achieves higher approximation ratios than \textit{RLVQC Global} across Maximum Cut and Minimum Vertex Cover instances, while is less competitive under Maximum Clique, indicating that restricting circuit synthesis to a block-based form is not in general detrimental and can be beneficial for some problems. Moreover, the circuits discovered by both RL variants frequently use substantially fewer $CX$ gates than standard QAOA, which is desirable in view of hardware constraints and noise sensitivity, while maintaining competitive solution quality. Second, we evaluate the learned blocks across problem sizes by learning blocks on $n=8$ instances and using them to build the ansatz required for problem instances on $n=12$ and $n=16$. The resulting solution quality remains stable for the larger problems, and statistical tests confirm that the improvements over QAOA and ma-QAOA baselines are robust across the tested configurations. The results also indicate that the learned blocks, when using a \texttt{Tied} parameter sharing strategy, achieve those solutions with a limited number of parameter optimization steps. These findings support the central premise that learned modular structure can be used to create effective ansatzes for larger problems, enabling ansatz design beyond the regime where ansatz search would be practical to run directly.

Overall, this work highlights the potential of reinforcement learning, even with simple agents and reward formulations, to discover useful circuit structures that can complement other known heuristics. At the same time, our goal is not to establish a new best solver for these optimization problems, but to validate a methodology for modular ansatz design which can be extended for larger problem instances. 

Future work may explore alternative composition rules, broader problem families, and more resource-aware objectives, for example by explicitly controlling circuit depth or two-qubit gate counts during block discovery. Another direction is to learn blocks under different training scenarios, \idest with different environment encodings or observations, without relying on a full simulation of the ansatz. Finally, it would be interesting to identify other tasks where modular ansatzes are beneficial and to study how the composition rule should be defined there, including how many modules to use, how many block types, and where to place them, for example in state preparation.
}

\bmhead{Acknowledgements}
We acknowledge the financial support from ICSC - ``National Research Centre in High Performance Computing, Big Data and Quantum Computing'', funded by European Union – NextGenerationEU. The project has been supported by the ESA Network of Resources Initiative.
We also acknowledge the support and computational resources provided by E4 Computer Engineering S.p.A.

\clearpage
\begin{appendices}

\section{Problem Description}
\label{app:problems_description}

In this appendix, we provide formal definitions of the combinatorial optimization problems used for our experiments: Maximum Cut, Maximum Clique, and Minimum Vertex Cover. For each problem, we present its mathematical formulation and the corresponding representation in QUBO form, which enables the application of quantum optimization algorithms.

\subsection{Maximum Cut Problem}

Let $G = (V, E)$ be an undirected graph, where $V$ is the set of $n$ vertices and $E$ is the set of edges. The Maximum Cut problem seeks a partition of $V$ into two disjoint subsets such that the number of edges between them is maximized.

This task can be formulated as a QUBO problem:
\begin{align}
Q(x) = \sum_{(i,j) \in E} x_i + x_j - 2x_i x_j,
\label{eq:_qubo}
\end{align}
where $x_i \in \{0,1\}$ is a binary variable indicating the side of the cut to which vertex $i$ is assigned.

\subsection{Minimum Vertex Cover Problem}

Given an undirected graph $G = (V, E)$, the Minimum Vertex Cover problem aims to find the smallest subset $S \subseteq V$ such that every edge $(i, j) \in E$ has at least one endpoint in $S$.

The corresponding QUBO formulation for a graph with $n$ vertices and penalty parameter $P > 0$ is:
\begin{align}
Q(x) = \sum_{i=1}^{n} x_i + P \sum_{(i,j) \in E} \left(1 - x_i - x_j + x_i x_j\right),
\label{eq:min_vertex_qubo}
\end{align}
where $x_i \in \{0,1\}$ indicates whether vertex $i$ is included in the cover (\idest $x_i = 1$).

\subsection{Maximum Clique Problem}

Given an undirected graph $G = (V, E)$, the Maximum Clique problem consists in finding the largest subset of vertices that form a fully connected subgraph, \idest a clique.
More formally, a subset $S \subseteq V$ is a clique if every pair of vertices in $S$ is connected by an edge, meaning the induced subgraph $G' = (S, E')$, where $E' = \{(i, j) \in E \mid i \in S,\, j \in S\}$, is fully connected.

The QUBO formulation for this problem is:
\begin{align}
Q(x) = -\sum_{i=1}^{n} x_i + P \sum_{(i,j) \in \overline{E}} x_i x_j,
\label{eq:MaxClique_QUBO}
\end{align}
where $x_i \in \{0,1\}$ indicates whether vertex $i$ is included in the clique, $\overline{E}$ denotes the set of non-edges in $G$, and $P > 0$ is a penalty parameter that discourages selecting non-adjacent vertices together.

\section{Implementation and Hyperparameter Configuration}
\label{app:implementation_hyperparameters}
\revision{
This appendix summarizes in Table~\ref{tab:impl_ppo_details} the implementation details needed to reproduce our experiments, including the PPO agent architecture and default hyperparameters. We implemented the agent based on OpenAI \texttt{Spinning Up} PPO implementation \footnote{The library is accessible here \url{https://github.com/openai/spinningup/tree/master}} based on PyTorch.
}

\begin{table*}[ht]
\centering
\footnotesize
\caption{Agent architecture and hyperparameters for RLVQC Block.}
\label{tab:impl_ppo_details}
\renewcommand{\arraystretch}{1.3}
\begin{tabular}{>{\raggedright\arraybackslash}m{6.5cm} l}
\toprule
\textbf{Component} & \textbf{Value} \\
\midrule
Policy/value networks       & Independent MLPs, 2 hidden layers of size 64 \\
Activation function         & Sigmoid \\
Weight initialization       & Kaiming uniform \\
\midrule
Discount factor $\gamma$        & 0.99 \\
GAE parameter $\lambda$         & 0.97 \\
Clipping parameter $\epsilon$   & 0.2 \\
Target KL                       & 0.01 \\
\bottomrule
\end{tabular}
\end{table*}

\section{Full Results for Experiment 1: Effectiveness of the Block Structure}
\label{app:effectiveness_block}

\begin{figure}[ht]
    \centering
    % Row 1
    \includegraphics[width=.49\textwidth]{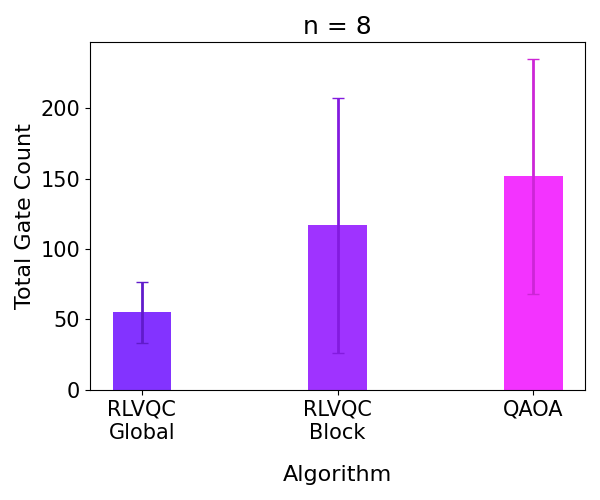}\hfill
    \includegraphics[width=.49\textwidth]{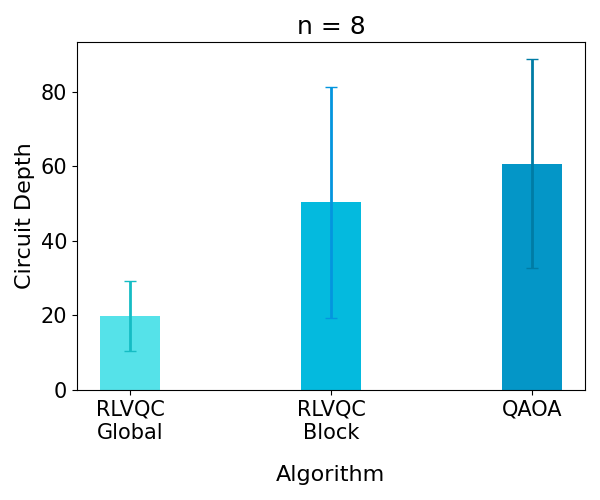}\\[\smallskipamount]
    
    % Row 2
    \includegraphics[width=.49\textwidth]{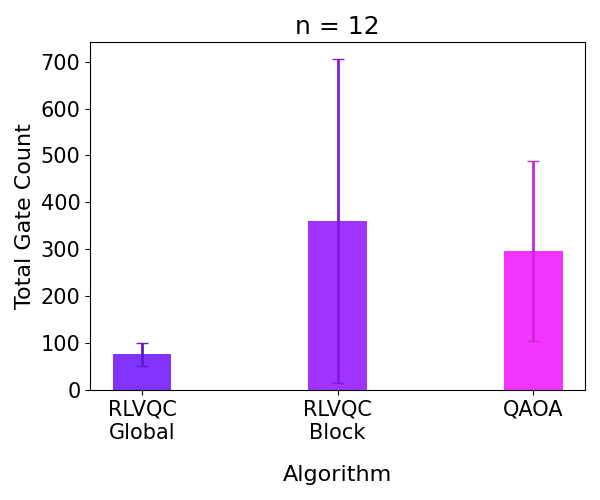}\hfill
    \includegraphics[width=.49\textwidth]{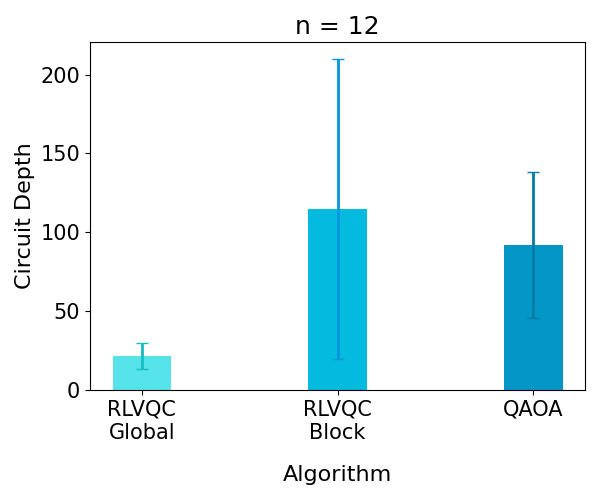}\\[\smallskipamount]
    
    \caption{Comparison of the average total gate count (left) and circuit depth (right) of the optimal circuits obtained using RLVQC Global, RLVQC Block, and the baseline QAOA, across instances with $n=8$ and $n=12$. Each histogram corresponds to a different algorithm. The bar above each histogram represents the standard deviation.}
    \label{fig:gate_count_and_depth_n=8_12}
\end{figure}

\begin{figure}[ht]
    \centering
    % First figure
    \includegraphics[width=0.88\textwidth]{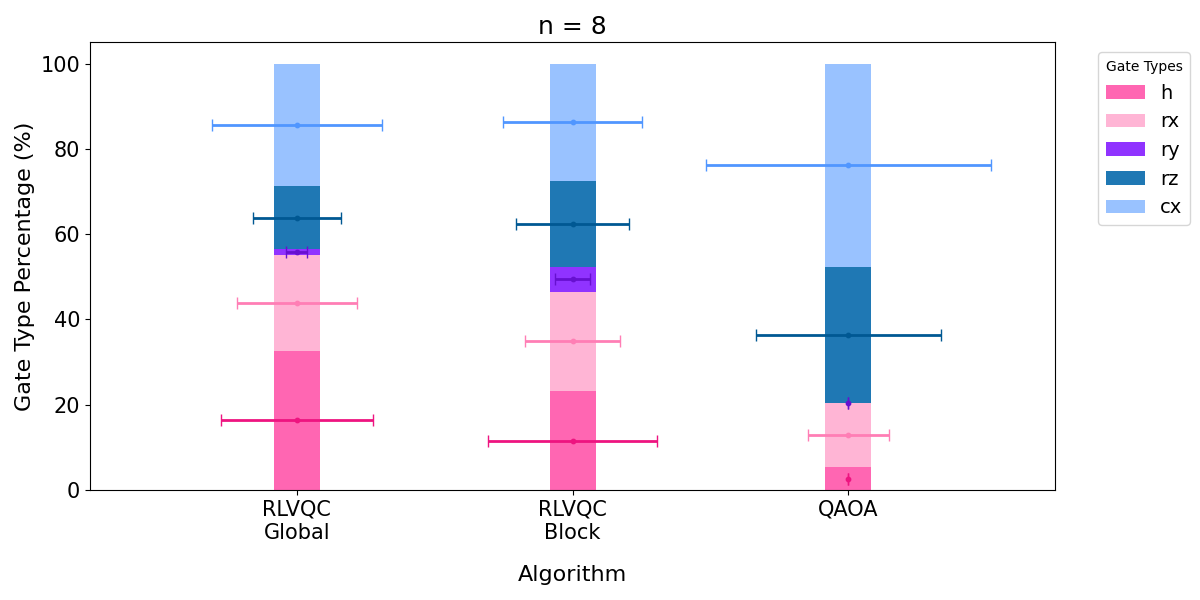}
    \vspace{1em} 
    % Second figure
    \includegraphics[width=0.88\textwidth]{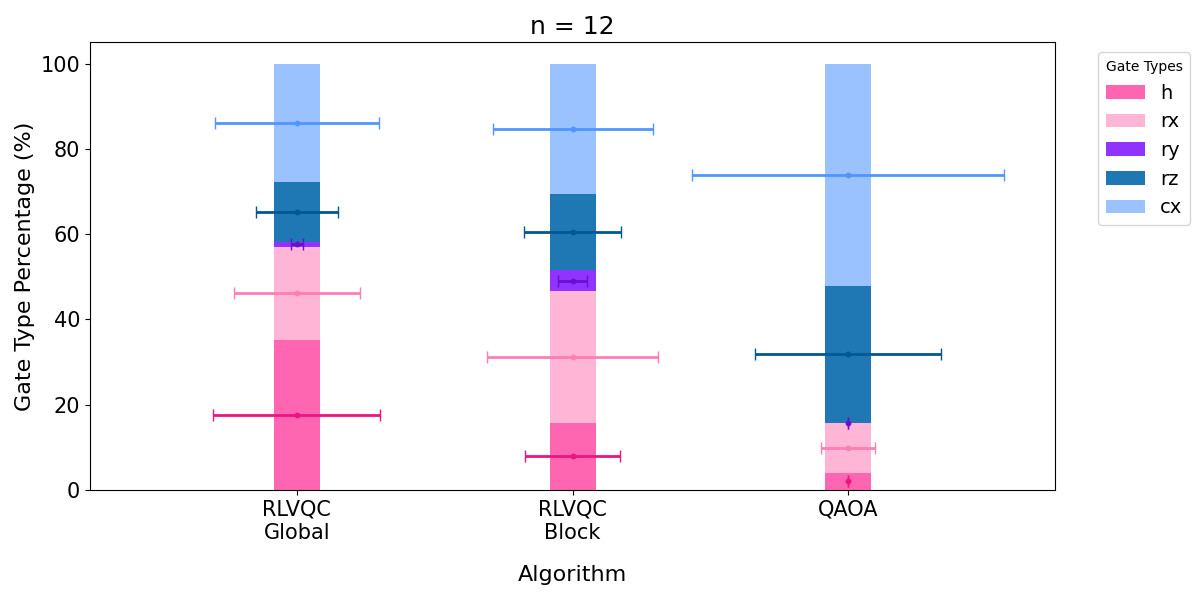}
    \vspace{1em}     \caption{
    Average gate usage by algorithm, expressed as a percentage of the total gate count in the optimal circuits produced by RLVQC Global, RLVQC Block, and the baseline QAOA, for instances with $n=8$ and $n=12$. Each histogram corresponds to a different algorithm. Standard deviation bars are proportional to the standard deviation within each algorithm.}
    \label{fig:gate_usage_qubo_n=8_12}
\end{figure}

\revision{
In this section, we report the complete results for the experiment on the effectiveness of the block structure, complementing the $n=16$ results presented in Section~\ref{sec:results:effectiveness_block}. Approximation ratios for $n=8$ and $n=12$ are reported in Tables~\ref{tab:ar_n=8} and~\ref{tab:ar_n=12}, respectively. The total gate count and circuit depth are shown in Fig.~\ref{fig:gate_count_and_depth_n=8_12}, while gate composition is reported in Fig.~\ref{fig:gate_usage_qubo_n=8_12}.
}

\begin{table*}[ht]
    \centering
    \footnotesize
    \caption{
    Approximation ratios achieved by RLVQC Global and RLVQC Block compared to the QAOA baseline on QUBO problem instances with $n = 8$ qubits, evaluated across various graph topologies. Each value reports the mean and standard deviation over five independent runs. For each instance, RLVQC results are typeset in \textbf{bold} when they outperform QAOA, and the highest average approximation ratio is \underline{underlined}.
    \label{tab:ar_n=8}
    }
    \resizebox{\textwidth}{!}{
    \begin{tabular}{cr|ccc}
    \toprule
        \textbf{Problem} & \textbf{Topology} & \textbf{QAOA} & \textbf{RLVQC Global} & \textbf{RLVQC Block} \\ 
        \midrule
        \multirow{8}{*}{\begin{tabular}{@{}c@{}}Maximum \\ Cut\end{tabular}} 
         & 2d-grid - 4 & 0.878 ± 0.032 & 0.715 ± 0.084 & \underline{\textbf{$\sim$ 1}} \\ 
        & 3-reg & 0.875 ± 0.058 & 0.806 ± 0.046 & \underline{\textbf{0.999 ± 0.001}} \\ 
        & barabási-albert - 2 & 0.878 ± 0.044 & 0.741 ± 0.057 & \underline{\textbf{0.999 ± 0.001}} \\ 
        & barabási-albert - 4 & 0.936 ± 0.037 & 0.829 ± 0.052 & \underline{\textbf{$\sim$ 1}} \\ 
        & cycle & 0.855 ± 0.012 & 0.679 ± 0.049 & \underline{\textbf{$\sim$ 1}} \\ 
        & erdős-rényi - 0.2 & 0.918 ± 0.031 & 0.759 ± 0.034 & \underline{\textbf{0.937 ± 0.013}} \\ 
        & erdős-rényi - 0.7 & 0.864 ± 0.057 & 0.789 ± 0.023 & \underline{\textbf{0.959 ± 0.003}} \\ 
        & star & 0.968 ± 0.021 & 0.710 ± 0.032 & \underline{\textbf{$\sim$ 1}} \\ 
        \midrule
        \multirow{8}{*}{\begin{tabular}{@{}c@{}}Maximum \\ Clique\end{tabular}}
        & 2d-grid - 4 & 0.801 ± 0.007 & \textbf{0.987 ± 0.004} & \underline{\textbf{0.995 ± 0.004}} \\ 
        & 3-reg & 0.977 ± 0.005 & \textbf{0.980 ± 0.011} & \underline{\textbf{0.986 ± 0.004}} \\ 
        & barabási-albert - 2 & 0.971 ± 0.015 & \textbf{0.982 ± 0.006} & \underline{\textbf{0.990 ± 0.006}} \\ 
        & barabási-albert - 4 & 0.966 ± 0.020 & \textbf{0.976 ± 0.008} & \underline{\textbf{0.979 ± 0.006}} \\ 
        & cycle & 0.917 ± 0.075 & \textbf{0.982 ± 0.004} & \underline{\textbf{0.993 ± 0.002}} \\ 
        & erdős-rényi - 0.2 & 0.823 ± 0.074 & \textbf{0.986 ± 0.004} & \underline{\textbf{0.991 ± 0.003}} \\ 
        & erdős-rényi - 0.7 & 0.920 ± 0.011 & \textbf{0.973 ± 0.012} & \underline{\textbf{0.984 ± 0.008}} \\ 
        & star & 0.855 ± 0.011 & \textbf{0.982 ± 0.012} & \underline{\textbf{0.993 ± 0.001}} \\ 
        \midrule
        \multirow{8}{*}{\begin{tabular}{@{}c@{}}Minimum \\ Vertex \\ Cover\end{tabular}}
        & 2d-grid - 4 & 0.956 ± 0.015 & \textbf{0.978 ± 0.005} & \underline{\textbf{$\sim$ 1}} \\ 
        & 3-reg & 0.973 ± 0.003 & \textbf{0.983 ± 0.001} & \underline{\textbf{0.996 ± 0.004}} \\ 
        & barabási-albert - 2 & 0.953 ± 0.014 & \textbf{0.978 ± 0.004} & \underline{\textbf{0.996 ± 0.004}} \\ 
        & barabási-albert - 4 & 0.920 ± 0.060 & \textbf{0.982 ± 0.006} & \underline{\textbf{0.986 ± 0.009}} \\ 
        & cycle & 0.967 ± 0.011 & \textbf{0.973 ± 0.009} & \underline{\textbf{$\sim$ 1}} \\ 
        & erdős-rényi - 0.2 & 0.944 ± 0.030 & \textbf{0.968 ± 0.009} & \underline{\textbf{0.988 ± 0.008}} \\ 
        & erdős-rényi - 0.7 & 0.967 ± 0.007 & \textbf{0.980 ± 0.007} & \underline{\textbf{0.997 ± 0.003}} \\ 
        & star & 0.912 ± 0.027 & \textbf{0.956 ± 0.014} & \underline{\textbf{0.994 ± 0.005}} \\      
    \bottomrule
    \end{tabular}
    }
\end{table*}

\begin{table*}[ht]
    \centering
    \footnotesize
    \caption{
    Approximation ratios achieved by RLVQC Global and RLVQC Block compared to the QAOA baseline on QUBO problem instances with $n = 12$ qubits, evaluated across various graph topologies. Each value reports the mean and standard deviation over five independent runs. For each instance, RLVQC results are typeset in \textbf{bold} when they outperform QAOA, and the highest average approximation ratio is \underline{underlined}.
    \label{tab:ar_n=12}
    }
    \resizebox{\textwidth}{!}{
    \begin{tabular}{cr|ccc}
    \toprule
        \textbf{Problem} & \textbf{Topology} & \textbf{QAOA} & \textbf{RLVQC Global} & \textbf{RLVQC Block} \\ 
        \midrule
        \multirow{8}{*}{\begin{tabular}{@{}c@{}}Maximum \\ Cut\end{tabular}} 
        & 2d-grid - 4 & 0.749 ± 0.093 & 0.599 ± 0.049 & \underline{\textbf{0.987 ± 0.025}} \\ 
        & 3-reg & 0.823 ± 0.025 & 0.666 ± 0.053 & \underline{\textbf{0.964 ± 0.054}} \\ 
        & barabási-albert - 3 & 0.828 ± 0.106 & 0.754 ± 0.061 & \underline{\textbf{0.995 ± 0.002}} \\ 
        & barabási-albert - 6 & 0.780 ± 0.034 & 0.766 ± 0.047 & \underline{\textbf{0.998 ± 0.001}} \\ 
        & cycle & 0.850 ± 0.016 & 0.596 ± 0.017 & \underline{\textbf{$\sim$ 1}} \\ 
        & erdős-rényi - 0.2 & 0.895 ± 0.046 & 0.691 ± 0.033 & \underline{\textbf{0.925 ± 0.003}} \\ 
        & erdős-rényi - 0.7 & 0.762 ± 0.018 & \textbf{0.817 ± 0.010} & \underline{\textbf{0.960 ± 0.019}} \\ 
        & star & 0.989 ± 0.013 & 0.642 ± 0.034 & \underline{\textbf{0.999 ± 0.001}} \\ 
        \midrule
        \multirow{8}{*}{\begin{tabular}{@{}c@{}}Maximum \\ Clique\end{tabular}}
        & 2d-grid - 4 & 0.770 ± 0.002 & \underline{\textbf{0.945 ± 0.003}} & \textbf{0.877 ± 0.072} \\ 
        & 3-reg & 0.771 ± 0.007 & \underline{\textbf{0.939 ± 0.006}} & \textbf{0.922 ± 0.054} \\ 
        & barabási-albert - 3 & 0.781 ± 0.011 & \textbf{0.941 ± 0.019} & \underline{\textbf{0.990 ± 0.003}} \\ 
        & barabási-albert - 6 & 0.826 ± 0.062 & \textbf{0.956 ± 0.013} & \underline{\textbf{0.979 ± 0.006}} \\ 
        & cycle & 0.779 ± 0.014 & \underline{\textbf{0.947 ± 0.013}} & \textbf{0.915 ± 0.043} \\ 
        & erdős-rényi - 0.2 & 0.768 ± 0.002 & \underline{\textbf{0.947 ± 0.018}} & \textbf{0.863 ± 0.068} \\ 
        & erdős-rényi - 0.7 & 0.878 ± 0.054 & \textbf{0.942 ± 0.010} & \underline{\textbf{0.986 ± 0.002}} \\ 
        & star & 0.772 ± 0.002 & \underline{\textbf{0.954 ± 0.008}} & \textbf{0.834 ± 0.069} \\ 
        \midrule
        \multirow{8}{*}{\begin{tabular}{@{}c@{}}Minimum \\ Vertex \\ Cover\end{tabular}}
        & 2d-grid - 4 & 0.961 ± 0.008 & \textbf{0.963 ± 0.011} & \underline{\textbf{0.997 ± 0.001}} \\ 
        & 3-reg & 0.974 ± 0.006 & 0.962 ± 0.008 & \underline{\textbf{0.993 ± 0.003}} \\ 
        & barabási-albert - 3 & 0.891 ± 0.066 & \textbf{0.977 ± 0.014} & \underline{\textbf{0.990 ± 0.002}} \\ 
        & barabási-albert - 6 & 0.802 ± 0.022 & \textbf{0.982 ± 0.009} & \underline{\textbf{0.990 ± 0.002}} \\ 
        & cycle & 0.961 ± 0.017 & 0.956 ± 0.012 & \underline{\textbf{0.999 ± 0.001}} \\ 
        & erdős-rényi - 0.2 & 0.970 ± 0.008 & 0.968 ± 0.005 & \underline{\textbf{0.987 ± 0.005}} \\ 
        & erdős-rényi - 0.7 & 0.804 ± 0.032 & \textbf{0.974 ± 0.012} & \underline{\textbf{0.997 ± 0.001}} \\ 
        & star & 0.934 ± 0.012 & \textbf{0.969 ± 0.004} & \underline{\textbf{0.999 ± 0.001}} \\     
    \bottomrule
    \end{tabular}
    }
\end{table*}

\end{appendices}

\FloatBarrier

\bibliography{bibliography}

\end{document}